\documentclass[fleqn,usenatbib]{mnras}

\usepackage{mathptmx}

\usepackage[T1]{fontenc}

\DeclareRobustCommand{\VAN}[3]{#2}
\let\VANthebibliography\thebibliography
\def\thebibliography{\DeclareRobustCommand{\VAN}[3]{##3}\VANthebibliography}

\usepackage{graphicx}	
\usepackage{amsmath}	
\usepackage{amssymb}	

\title[Spectral features of blazars]{The optical spectral features of 27 $Fermi$ blazars}

\author[B. K. Zhang et al.]{
Bing-Kai Zhang,$^{1,2}$\thanks{E-mail: zhangbk\_ynu@163.com}
Wei-Feng Tang,$^{1}$
Chun-Xiao Wang,$^{1,2}$
Qi Wu,$^{1,2}$\thanks{E-mail: cubewuqi@163.com}
\newauthor
Min Jin,$^{1,3}$
Ben-Zhong Dai$^{4,5}$
and Feng-Rong Zhu$^{3}$
\\
$^{1}$Department of Physics, Fuyang Normal University, Fuyang 236037, China\\
$^{2}$Key Laboratory of Functional Materials and Devices for Informatics
      Anhui Higher Education Institutes, Fuyang 236037, China\\
$^{3}$Department of Applied Physics, Southwest Jiaotong University, Chengdu 610031, China\\
$^{4}$Key Laboratory of Astroparticle Physics, Yunnan Province, Kunming 650091, China\\
$^{5}$Department of Physics, Yunnan University, Kunming 650091, China
}

\date{Accepted 2022 December 20. Received 2022 December 1; in original form 2022 October 21}

\pubyear{2022}

\graphicspath{{./}{mfigures/}}

\begin{document}
\label{firstpage}
\pagerange{\pageref{firstpage}--\pageref{lastpage}}
\maketitle

\begin{abstract}
A spectral variation accompanied with flux variability is a commonly-observed phenomenon for blazars.
In order to further investigate the optical spectral feature of blazars, we have collected the long-term
optical $V$ and $R$ band data of 27 blazars (14 BL Lacs and 13 FSRQs), and calculated their optical spectral indices.
The results show that the spectral indices vary with respect to the brightness for all of these blazars.
In general, the optical spectrum progressively becomes flatter (or steeper), when the brightness increases.
However the spectrum changes more and more slowly, until it tends to be stable.
In other words, the source becomes bluer (or redder) and then gradually stabilizes when it brightens,
which are briefly named the bluer-stable-when-brighter (BSWB) and redder-stable-when-brighter (RSWB) behaviors, respectively.
Thirteen of the 14 BL Lacs show the BSWB behavior, with an exception of AO 0235+164. On the contrary, most of FSRQs (10 out of 13) exhibit the RSWB trend.
It is confirmed that blazars follow the two universal optical spectral behaviors, namely, BSWB and RSWB.
The model of two constant-spectral-index components can well explain the optical spectral features qualitatively and quantitatively.
The results illustrate that the optical emission are mainly composed of two stable-color components, i.e., less variable thermal emission and high variable synchrotron radiation.
And in most cases, the thermal component of BL Lacs is redder than that of synchrotron radiation, whereas FSRQs are the opposite.
\end{abstract}




\section{Introduction}

Blazars are a subclass of the active galactic nuclei (AGN).  According to the optical emission/absorption line features, they can be divided into two subclasses: flat spectrum radio quasars (FSRQs) and BL Lacaerte objects (BL Lacs). Blazars have been observed at all wavebands from radio to gamma-ray bands. Their spectral energy distributions display a typical structure with two main peaks which locate at the low energy bands from the radio to the X-ray band, and high energy gamma-ray band, respectively \citep[e.g.][]{falomo14}.

It is widely accepted that the low energy component arises mainly from synchrotron radiation emitted by ultrarelativistic electrons in the jet.
In addition, the accretion disk, dusty torus and broad line region (BLR) have also significant contributions to the low energy component \citep[e.g.][]{ghisellini19}.
The high energy peak can be explained with leptonic models, in which the high energy $\gamma-$rays are generated by inverse Compton scattering of the low-energy photons by the same ultrarelativistic electrons in the jet \citep[e.g.][]{dermer93,sokolov04,bottcher13} ,
or with hadronic models, in which the high-energy emission is dominated by the synchrotron radiation of ultrarelativistic
protons or the photopion production followed by pion decay and subsequent cascade \citep[e.g.][]{mucke03,murase12}.
However, the emission mechanism remains an open question.

Blazars exhibit large and rapid variations on a variety of time scales from years to even shorter than an hour, and their flux variations are always accompanied by their color variations, i.e., optical spectral variations.
The study of the color variations is helpful to understand the emission mechanisms. Many works have investigated the color variation behaviors in the optical band \citep[e.g.][and references therein]{vagnetti03,gu06,rani10,ikejiri11,bonning12,wierzcholska15,zhang15,mao16,li18,meng18,gaur19,safna20,negi22,zhang22}. Generally, there are two main types of color variation behaviors. One behavior is commonly called bluer-when-brighter (BWB) in brief, which means that the source becomes bluer (i.e., spectrum becomes flatter) when the brightness increases \citep[e.g.][]{takalo89,raiteri15,dai09,gupta19}. Another phenomena is generally referred to as redder-when-brighter (RWB), which denotes that the source turns redder (i.e., spectrum turns steeper) as the brightness increases \citep[e.g.][]{villata06,raiteri17}.

Previous results showed that most of FSRQs exhibit redder-when-brighter trends, and BL Lacs tend to display bluer-when-brighter hehaviors.
However, there also exists other special phenomena, and there seem to be lack of unique behaviors for blazars.
Meanwhile, BWB and RWB trends cannot always well describe the characteristics of the color variations of blazars \cite[e.g.][]{zhang15}.
Therefore, the segmented linear functions were attempted to characterize the color variation features \cite[e.g.][]{yuan17,fan18,sarkar19,safna20,xiong20,fang22}. However, it is difficult to explain the abrupt inflection.
Recently, \cite{zhang22} have found two universal phenomena universal optical spectral behaviors for blazars with the Small and Moderate Aperture Research Telescope System (SMARTS) monitoring data.
In the low state, the optical spectrum gradually becomes flatter or steeper, but more and more slowly as the brightness increases, and then tends to stabilize in the high state. These two phenomena are briefly named the bluer-stable-when-brighter (BSWB) and redder-stable-when-brighter (RSWB) behaviors, respectively. They have also successfully constructed a nonlinear function to describe quantitatively the optical behavior of blazars well, and given a reasonable physical explanation.

The purpose in this work is to further investigate the optical spectral features with a different large sample of fermi  blazars. In Section 2, the blazar sample and their light curves of optical $V$ and $R$ bands are introduced. And then, the optical spectral indices are calculated and the spectral behaviors are described in Section 3, followed by a detailed discussion in Section 4 and brief conclusions in the last section.

\begin{figure}
	\includegraphics[width=1.0\columnwidth]{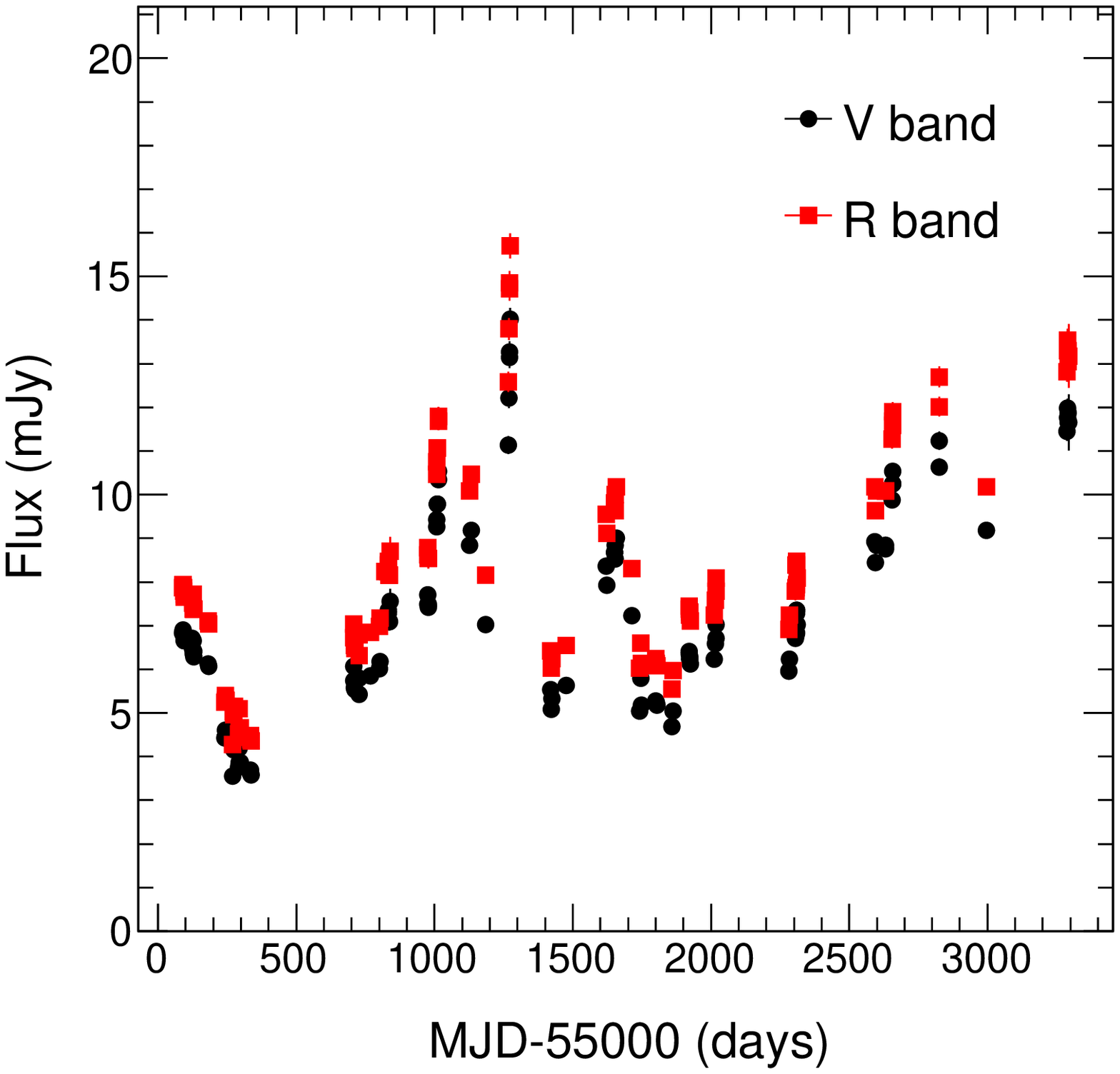}
    \includegraphics[width=1.0\columnwidth]{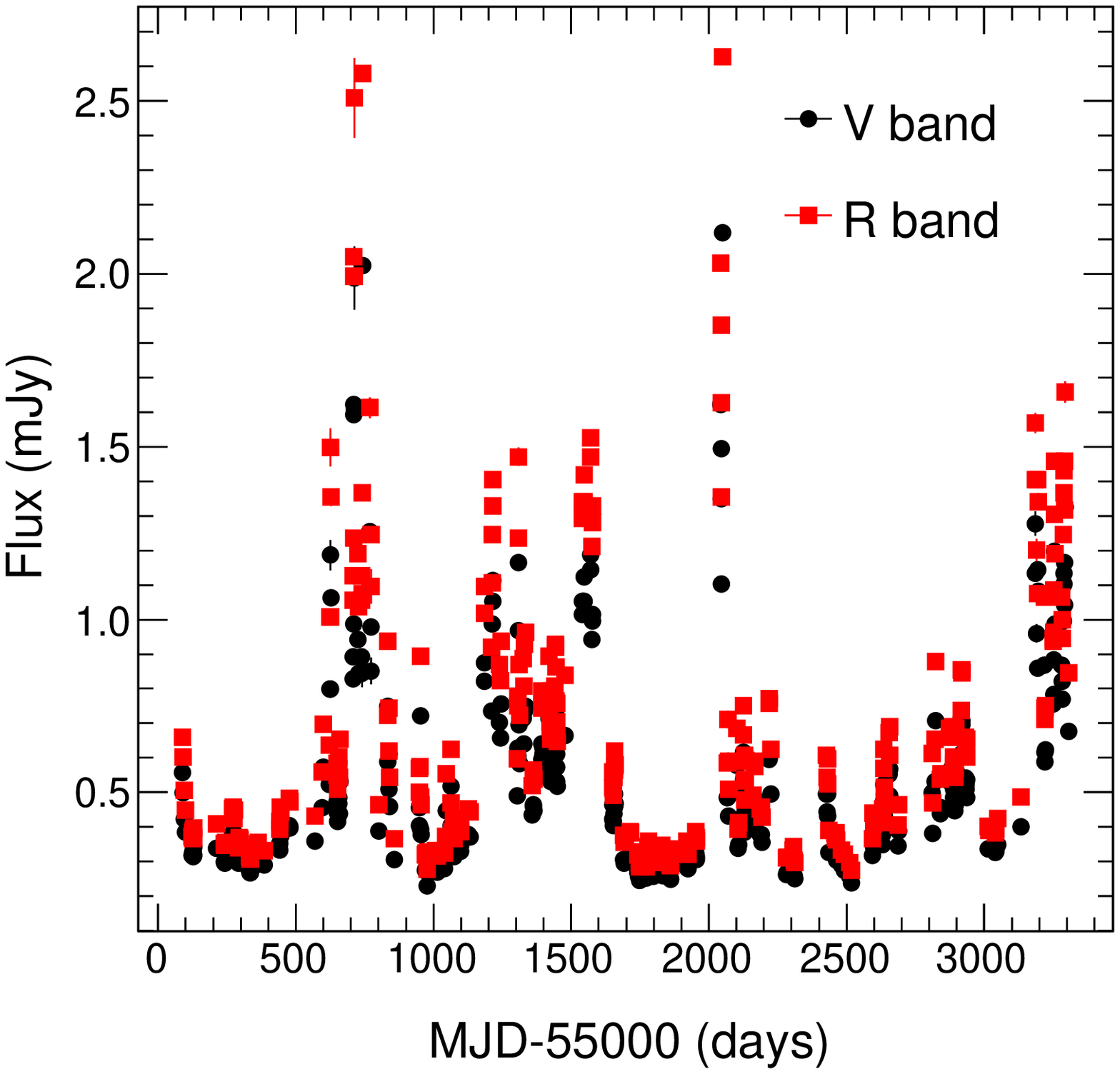}
    \caption{The optical $V$ and $R$ band light curves for BL Lac object 1ES 1959+650 (upper panel) and FSRQ B2 1633+382 (lower panel).}
    \label{lightcurve}
\end{figure}

\begin{figure*}
	\includegraphics[width=\columnwidth]{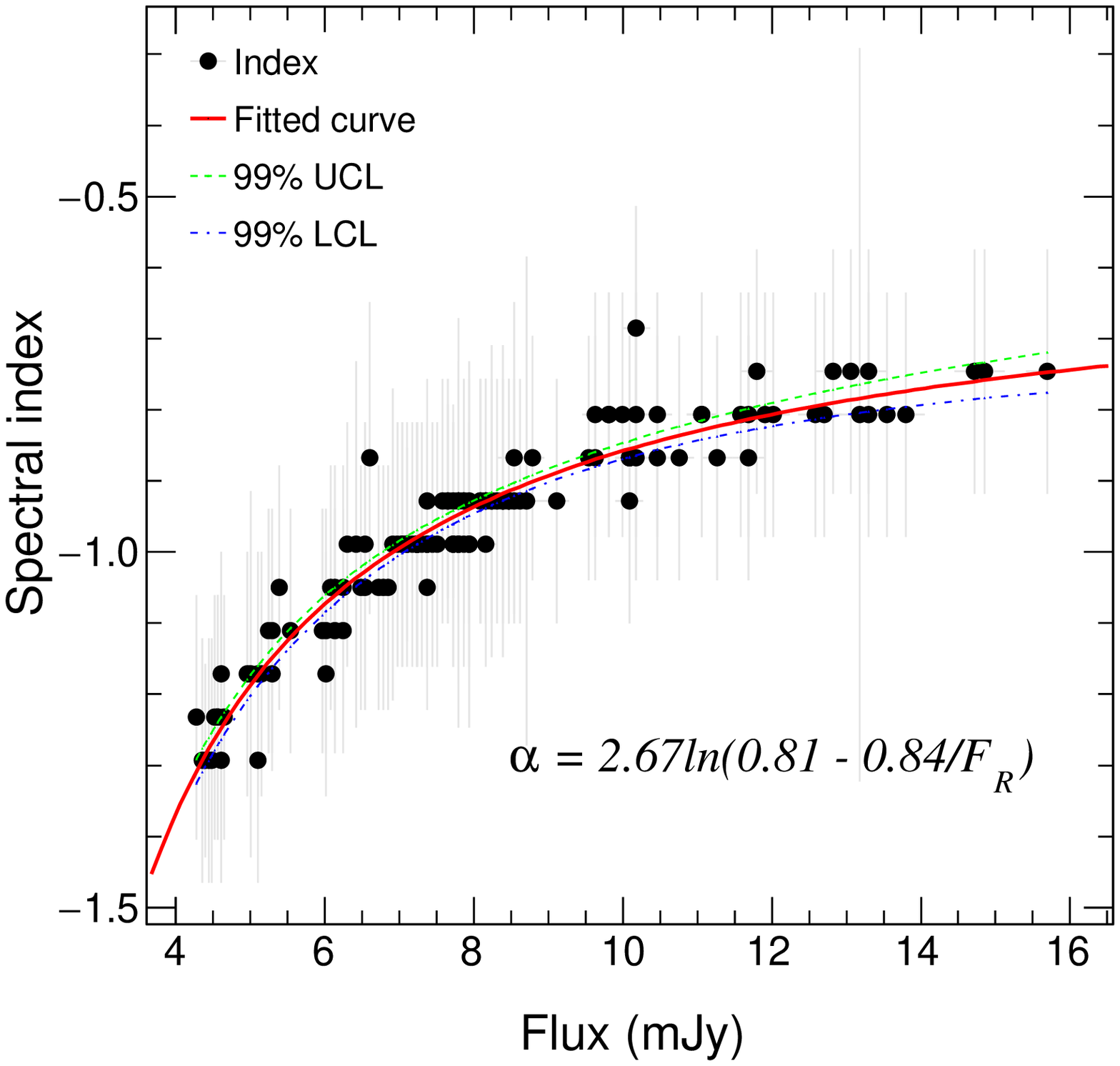}
    \includegraphics[width=1.0\columnwidth]{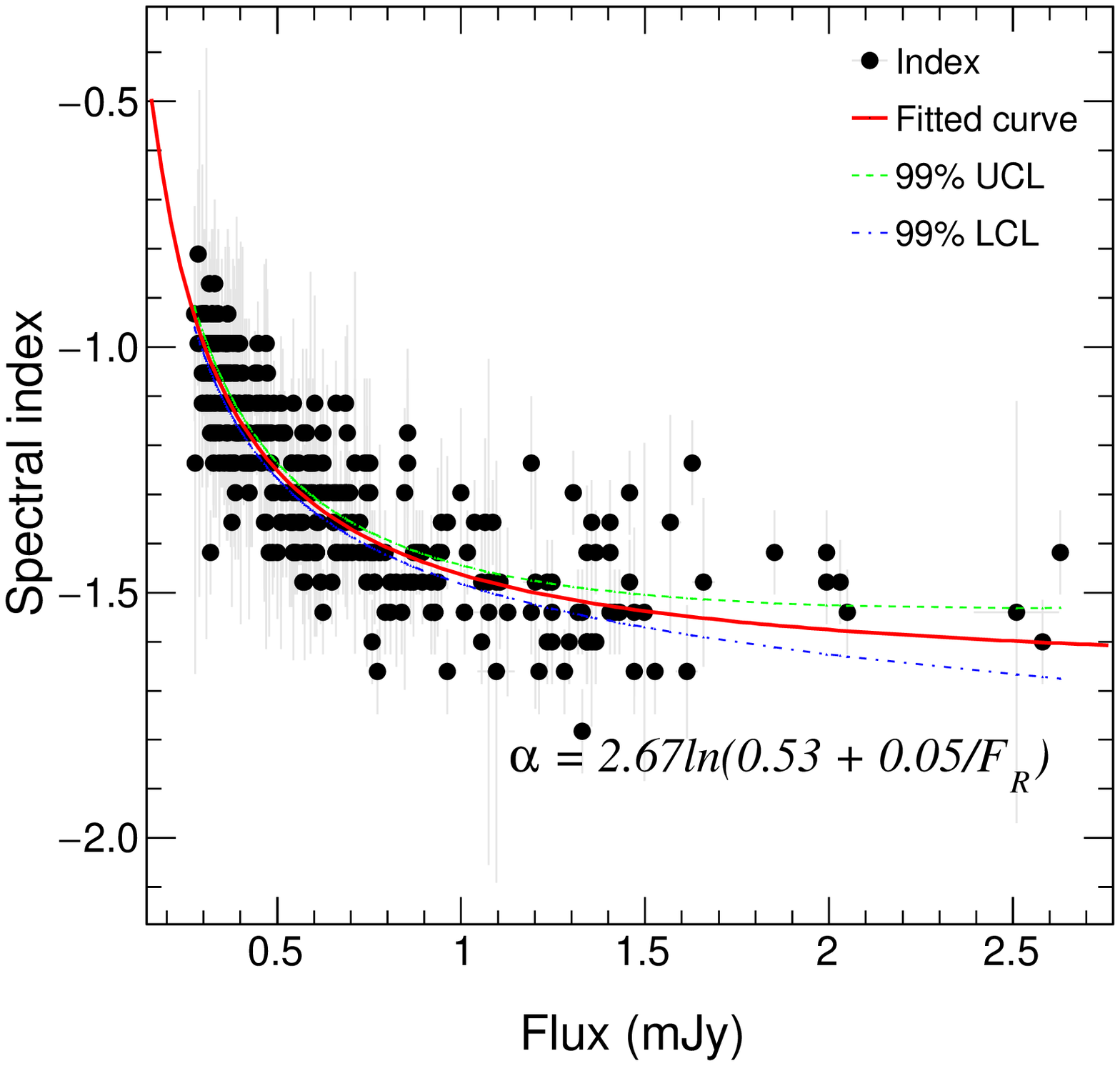}
    \caption{The optical spectral index against the flux of optical $R$ band (black dots) for BL Lac object 1ES 1959+650 (left panel) and FSRQ B2 1633+382 (right panel). The red solid line indicates the result of spectral behavior fitted with equation~(\ref{Equ:fitln}). The green dashed and blue dash-dotted lines mean the 99\% confidence level bands of the fitted curve.
    Note that due to the observation resolution, almost all the dots are distributed on horizontal lines with different values, and some of the dots overlap together.}
    \label{fig:spec-example}
\end{figure*}

\begin{table}
	\centering
	\caption{The Characteristics of 27 Fermi Blazars.}
	\label{tab:spec}
	\begin{tabular}{llrcc} %
		\hline
		Source & Class & N & A$_{V}$ & Behavior\\
        \hline
     3C 66A 	    &BL Lac  & 352 & 0.231  & BSWB\\
     AO 0235+164    &BL Lac  & 182 & 0.219  & RSWB\\
     S5 0716+714    &BL Lac  & 206 & 0.085  & BSWB\\
     PKS 0735+178   &BL Lac  &  34 & 0.094  & BSWB\\
     OJ 287         &BL Lac  & 491 & 0.077  & BSWB\\
     Mrk 421        &BL Lac  & 555 & 0.042  & BSWB\\
     H 1219+305     &BL Lac  &  14 & 0.056  & BSWB\\
     W Com 	        &BL Lac  & 315 & 0.064  & BSWB\\
     H 1426+428     &BL Lac  &  14 & 0.034  & BSWB\\
     Mrk 501        &BL Lac  & 448 & 0.052  & BSWB\\
     1ES 1959+650   &BL Lac  & 131 & 0.474  & BSWB\\
     PKS 2155-304   &BL Lac  & 296 & 0.060  & BSWB\\
     BL Lac         &BL Lac  & 672 & 0.901  & BSWB\\
     1ES 2344+514   &BL Lac  &  80 & 0.580  & BSWB\\
     \hline
     CTA 26         &FSRQ    &  24 & 0.253  & RSWB\\
     PKS 0420-014   &FSRQ    &  51 & 0.345  & RSWB\\
     PKS 0736+01    &FSRQ    & 130 & 0.375  & SWB \\
     OJ 248         &FSRQ    & 193 & 0.092  & RSWB\\
     Ton 599        &FSRQ    & 198 & 0.054  & RSWB\\
     PKS 1222+216   &FSRQ    & 422 & 0.063  & RSWB\\
     3C 273         &FSRQ    & 277 & 0.057  & BSWB\\
     3C 279 	    &FSRQ    & 479 & 0.078  & BSWB\\
     PKS 1510-08    &FSRQ    & 352 & 0.275  & RSWB\\
     B2  1633+382   &FSRQ    & 340 & 0.033  & RSWB\\
     3C 345         &FSRQ    &  41 & 0.036  & RSWB\\
     CTA 102        &FSRQ    & 304 & 0.198  & RSWB\\
     3C 454.3       &FSRQ    & 531 & 0.289  & RSWB\\
		\hline
	\end{tabular}
\end{table}

\section{Data Collection}

We have collected the optical $V$ and $R$ band photometry data from the long-term monitoring program carried out at Steward Observatory, University of Arizona. The observations were performed from 2008 to 2018 by the 1.54-m Kuiper and the 2.3-m Bok telescopes, and all of the observed data are publicly available at the website\footnote{http://james.as.arizona.edu/$\sim$psmith/Fermi}. Details of the observation and data analysis were described by \cite{smith09}.

In order to have enough data points and ensure the quality of the spectral behaviors of blazars, we only select the object which has at least 10 pairs of quasi-simultaneous $VR$ photometry observations. Thus, 27 blazars survive, which consist of 14 BL Lacs and 13 FSRQs. Most of them have hundreds of pairs of $VR$ photometry observations, and only a few sources have less than 100 pairs. These sources as well as their optical types and the numbers of observation pairs are given in Columns 1 - 3 of Table~\ref{tab:spec}, respectively.

The magnitudes of each source are converted into flux densities after the Galactic interstellar reddening correction. The values of Galactic extinction, $A_{\lambda}$, of $V$ band are taken from the NASA Extragalactic Database (NED\footnote{https://ned.ipac.caltech.edu/}), and are listed in Column 4 of Table~\ref{tab:spec}. Fig.~\ref{lightcurve} gives two examples, which illustrates the optical $VR$ light curves for BL Lac object 1ES 1959+650 and FSRQ B2 1633+382. The light curves show violent and large amplitude variations, and exhibit the similar variation trends between $V$ and $R$ bands.

\section{Optical Spectral behavior}

Since the optical spectrum follows a power law and there are only two optical band ($V$ and $R$) measurements, we have determined the spectral index by the formula
\begin{equation}
   \alpha = \frac{logF_{\nu_{2}}-logF_{\nu_{1}}}{log\nu_{2} - log\nu_{1}},
   \label{eq:lnfit}
\end{equation}
where $\nu$ represents the frequency, and $F_{\nu}$ represents the flux. The spectral index for each $V$ and $R$ pair has been obtained.

\begin{figure*}
	\includegraphics[width=1.0\columnwidth]{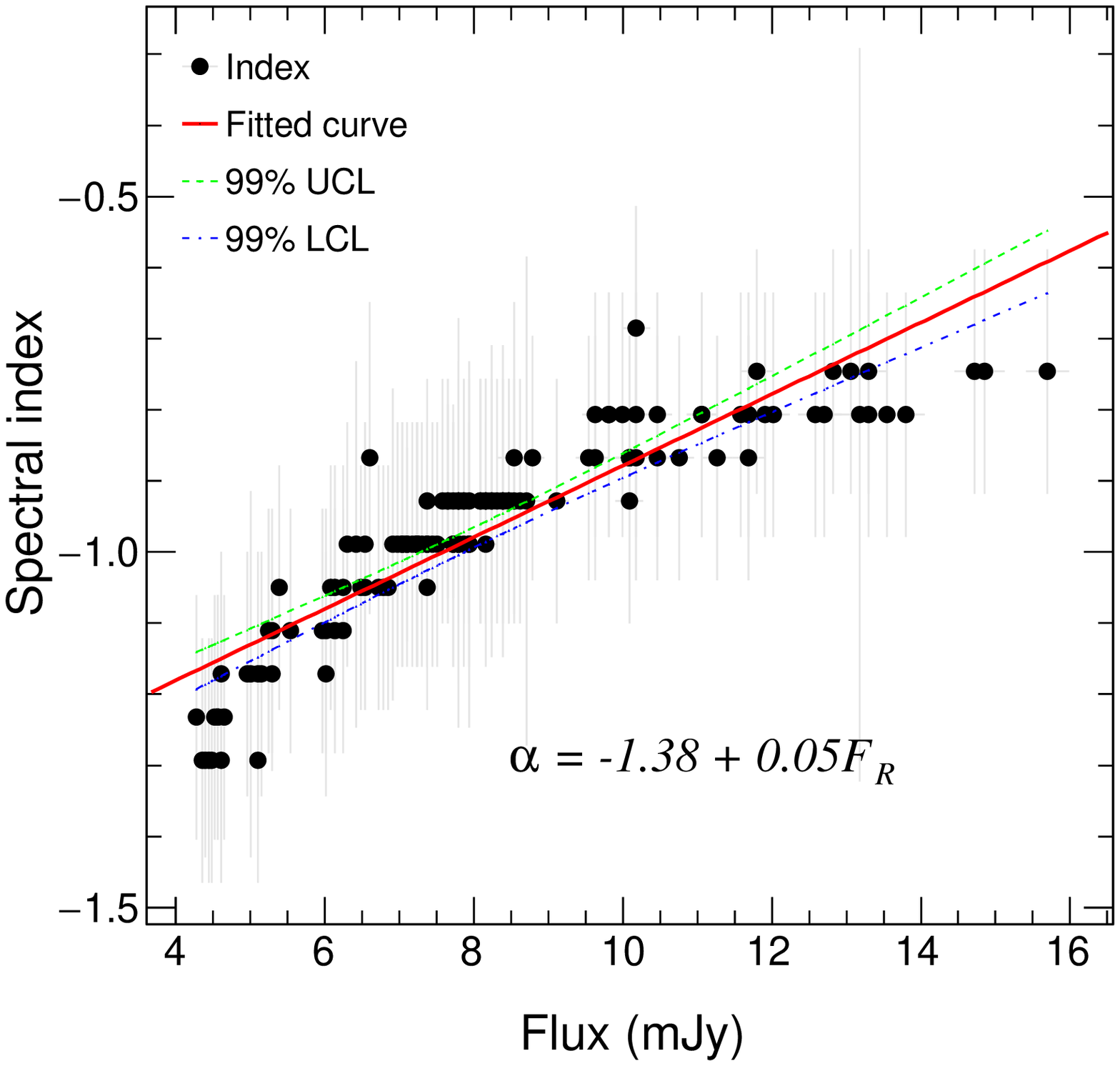}
    \includegraphics[width=1.0\columnwidth]{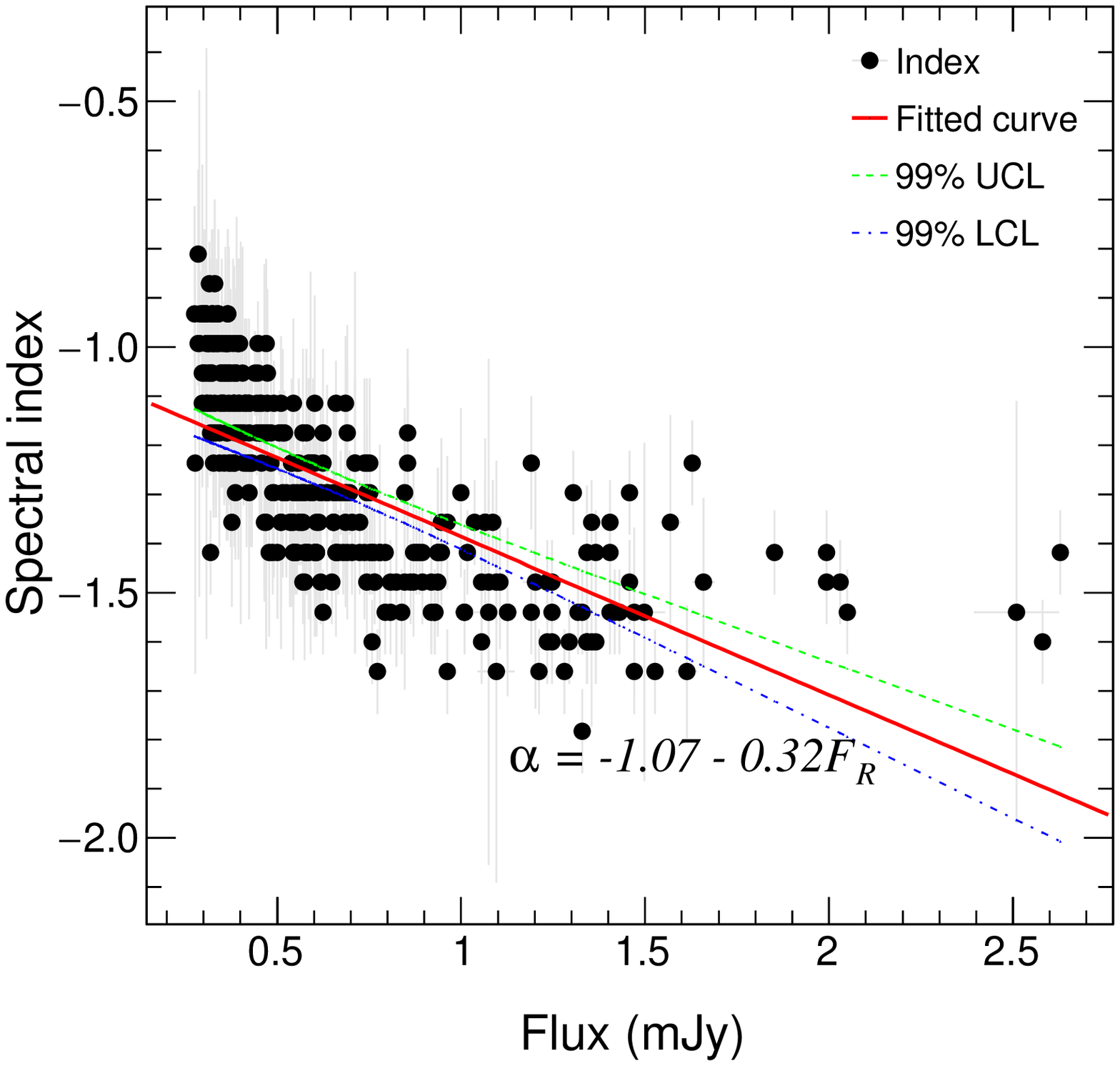}
    \caption{The same as Fig.~\ref{fig:spec-example}, except that equation~(\ref{Equ:fitpol1}) is used instead of equation~(\ref{Equ:fitln}).}
    \label{fig:spec-pol1}
\end{figure*}

The relation of the optical spectral index $\alpha$ with the flux $F_{R}$ has been explored.
As a typical example, the left panel of Fig.~\ref{fig:spec-example} plots the spectral index versus optical $R$ band flux of BL Lac object 1ES 1959+650, which shows spectral index varies with the flux.
From the figure, one can distinctly see the value of the optical spectral index gradually gets larger as the flux increases, which means that the optical spectrum becomes flatter (harder) when the flux increases, in the sense that the source becomes bluer when it brightens.
However, the spectral index and the flux follow a distinct non-linear relation. It is clear that the spectral index increases step by step as the flux increases, but more and more slowly, and finally tends to be stable. This phenomenon is the typical bluer-stable-when-brighter (BSWB) trend.

The right panel of Fig.~\ref{fig:spec-example} displays another representative example, an opposite spectral behavior of FSRQ B2 1633+382. When the source brightens, the value of the spectral index decreases gradually, i.e. the optical spectrum progressively becomes steeper (softer), but more and more slowly, and then eventually tends to stabilize. In other words, the color becomes redder, and tends to be stable when the source brightens. This behavior is the representative redder-stable-when-brighter (RSWB) trend.

Although the behaviors of BSWB and RSWB exhibit opposite trends in low states, they share a common feature that the spectrum tend to stabilize when the source becomes brighter. To quantitatively characterize the relationship between the spectral index $\alpha$ and the $R$ band flux $F_{R}$,  we employ a non-linear function,
\begin{equation}
\alpha = 2.67ln(a+b/F_{R}),
\label{Equ:fitln}
\end{equation}
proposed by \cite{zhang22}, to fit the curve of $\alpha$ versus $F_{R}$, with two free parameters of $a$ and $b$. The fitted results as well as the 99\% confidence level bands are superimposed on Fig.~\ref{fig:spec-example} with dashed and dash-dotted lines.
The $\chi^{2}$ values are 6.58 and 315.69, respectively.
Both of them are consistent well with the trends of the spectral behaviors.

To compare with the linear fitting, we fit the curves of Fig.~\ref{fig:spec-example} with the equation
\begin{equation}
\alpha = a+bF_{R},
\label{Equ:fitpol1}
\end{equation}
with two free parameters of $a$ and $b$. The results are plotted on Fig.~\ref{fig:spec-pol1}.
The $\chi^{2}$ values are 16.77 and 599.27, respectively.
They do not seem to be as good as those in Fig.~\ref{fig:spec-example}.
We carry out White's test to diagnose whether there is heteroscedasticity in the fitting process. For equation~(\ref{Equ:fitpol1}), White's test gives $p$-values are 3.89$\times10^{-7}$ and 1.0$\times10^{-12}$ for 1ES 1959+650 and B2 1633+382, respectively. For equation~(\ref{Equ:fitln}), the corresponding $p$-values are 0.74 and 0.25. The test means that at 1\% significance level, the fitting with equation~(\ref{Equ:fitpol1}) exist evidence heteroscedasticity, while the fitting with equation~(\ref{Equ:fitln}) has no heteroscedasticity. White's test suggests that the fitting results with non-linear equation~(\ref{Equ:fitln}) are better than those with linear equation~(\ref{Equ:fitpol1}).


For two equations, we also calculate the Bayesian information criterion (BICs) which is one of the popular statistical approaches for the comparative evaluation among different models. The model with the lower BIC values is preferred. For these two blazars, BICs of equation~(\ref{Equ:fitln}) are 10 and 284 lower than those of equation~(\ref{Equ:fitpol1}), which suggests that equation~(\ref{Equ:fitln}) is the preferred model to describe the spectral behavior.

\begin{figure}
    \includegraphics[width=1.0\columnwidth]{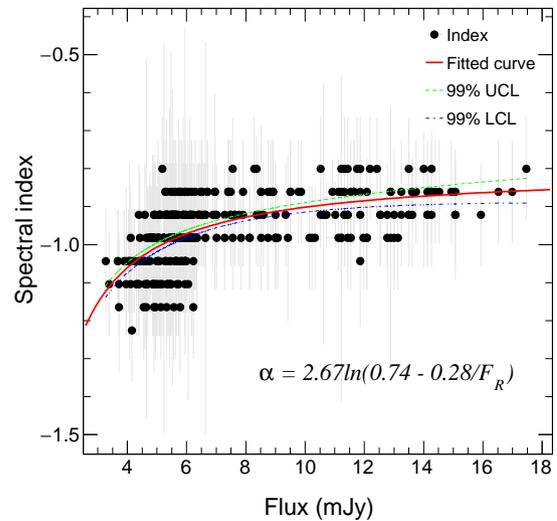}
    \caption{Same as Fig.~\ref{fig:spec-example}, but for BL Lac object 3C 66A.
    }
    \label{fig:spec-bllac0}
\end{figure}

\begin{figure*}
    \includegraphics[width=0.68\columnwidth]{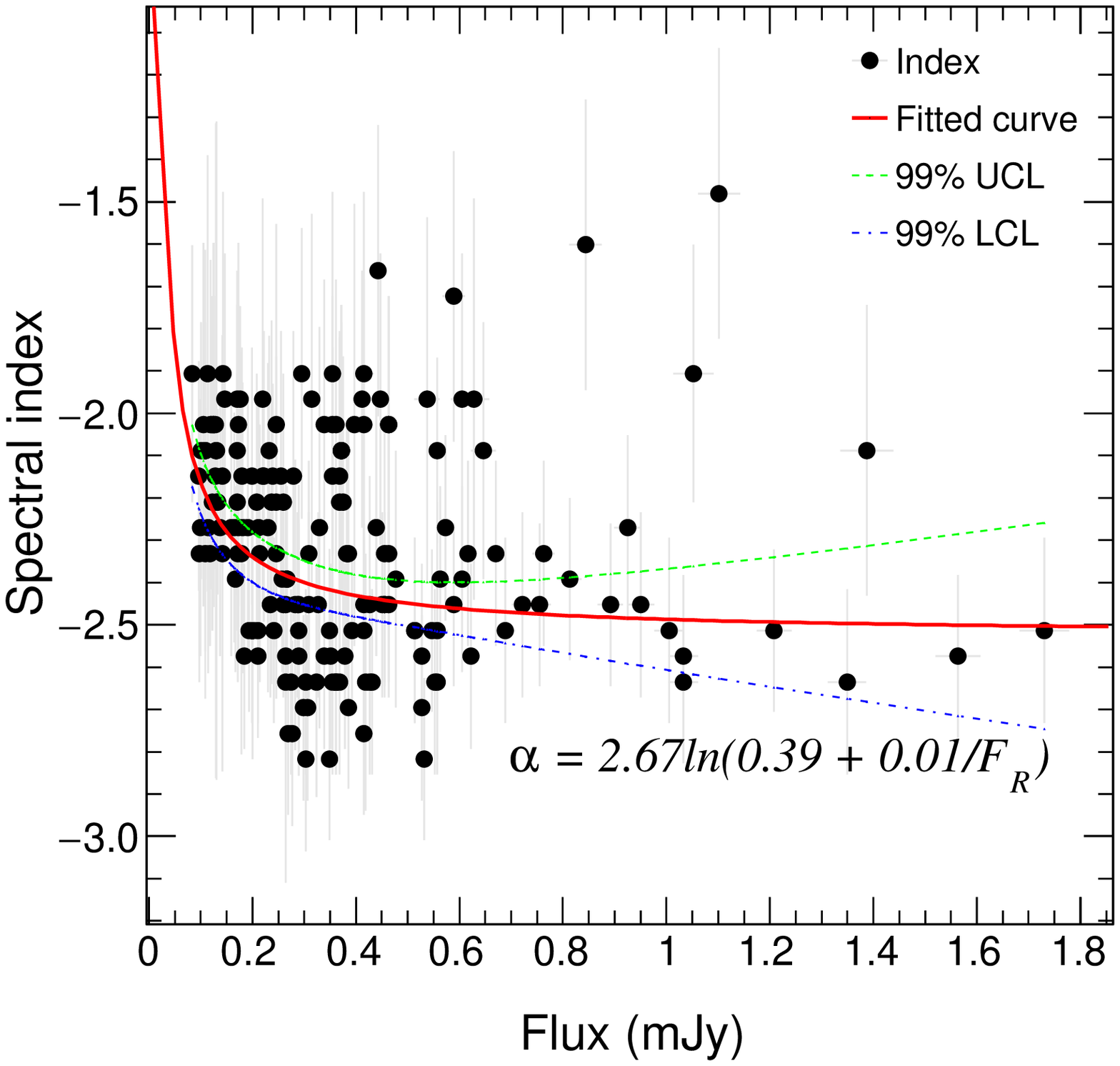}
    \includegraphics[width=0.68\columnwidth]{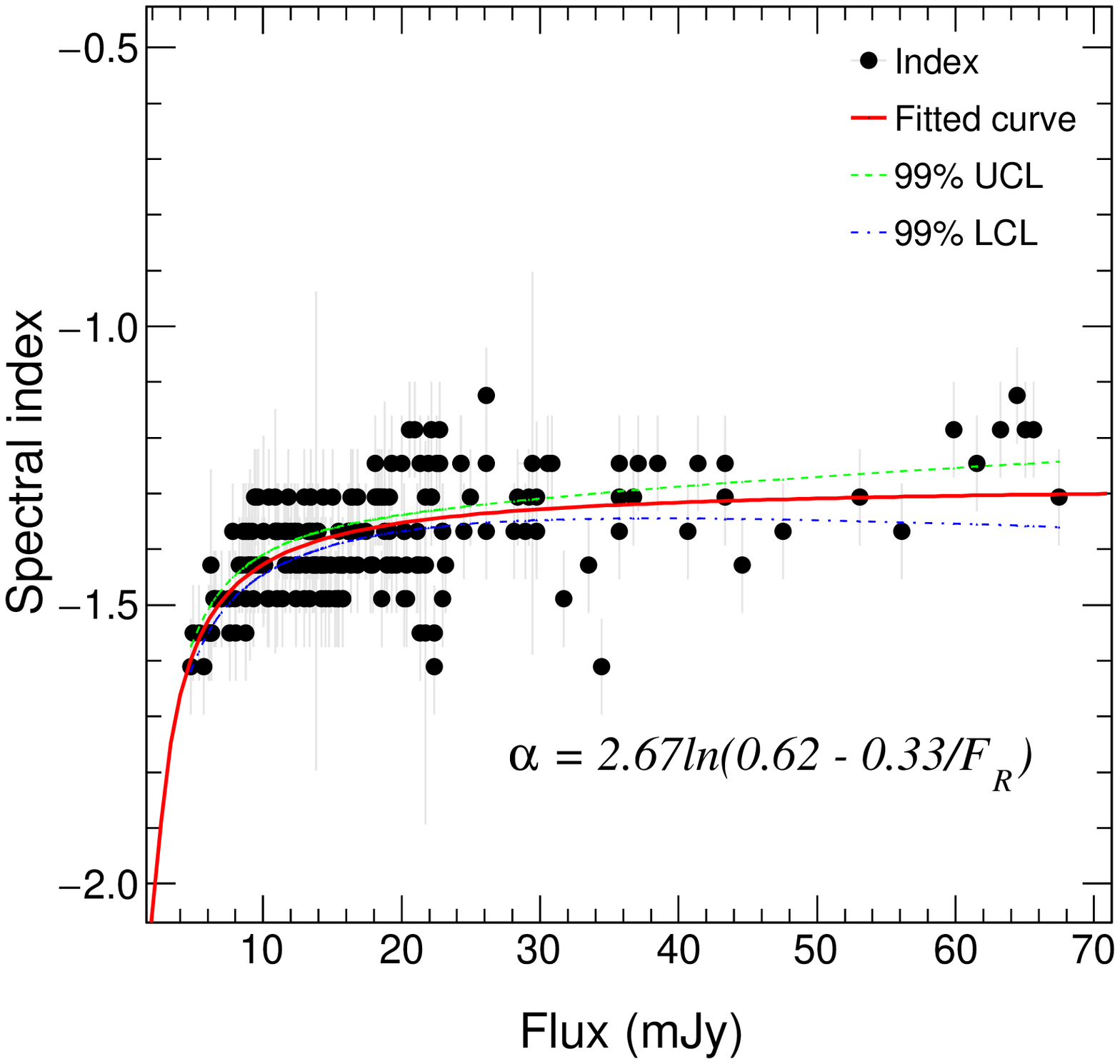}
    \includegraphics[width=0.68\columnwidth]{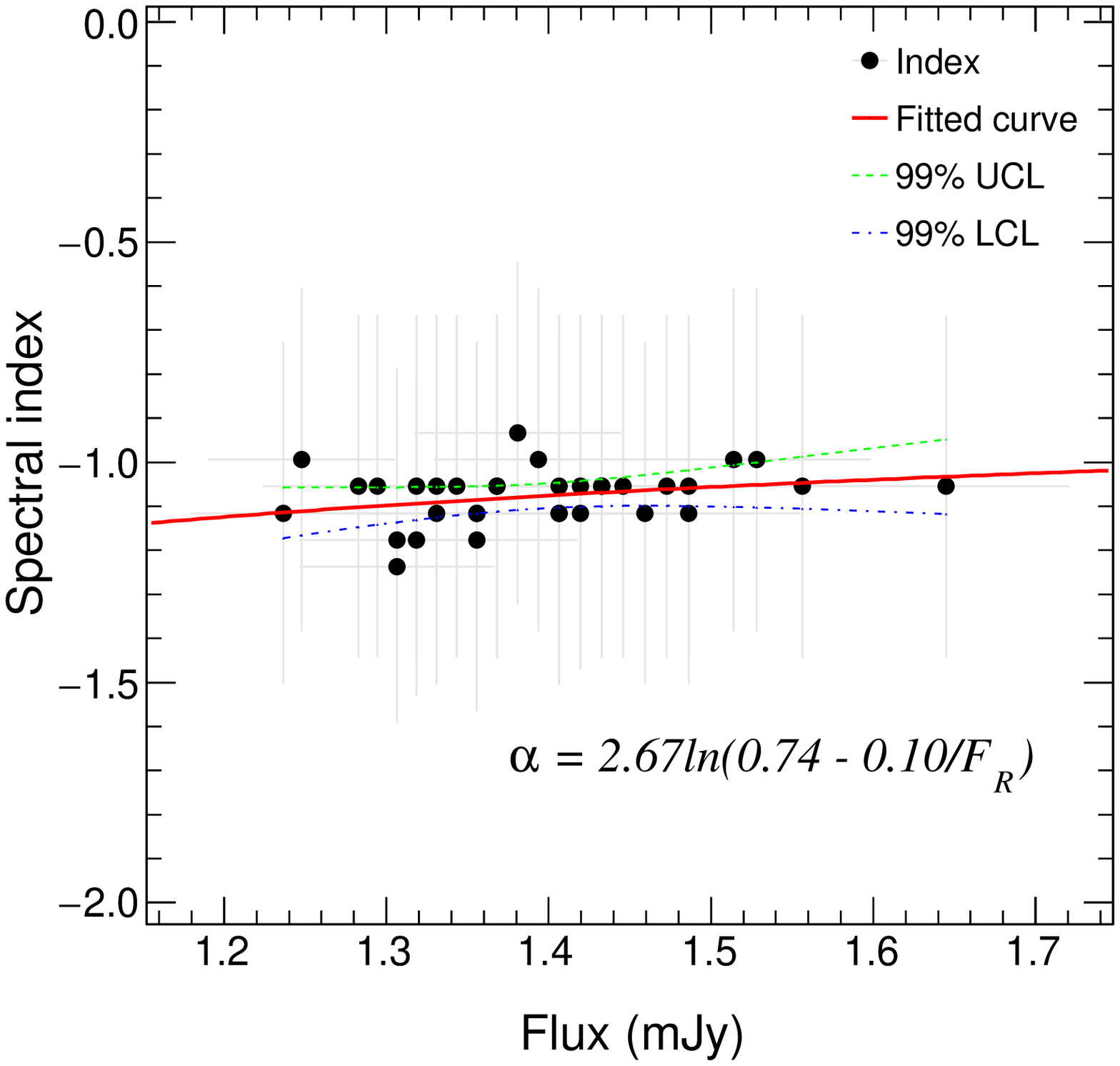}
    \caption{Same as Fig.~\ref{fig:spec-example}, but for BL Lac objects (a) AO 0235+164, (b) S5 0716+714 and (c) PKS 0735+178 from left to right.}
    \label{fig:spec-bllac1}
\end{figure*}

\begin{figure*}
    \includegraphics[width=0.68\columnwidth]{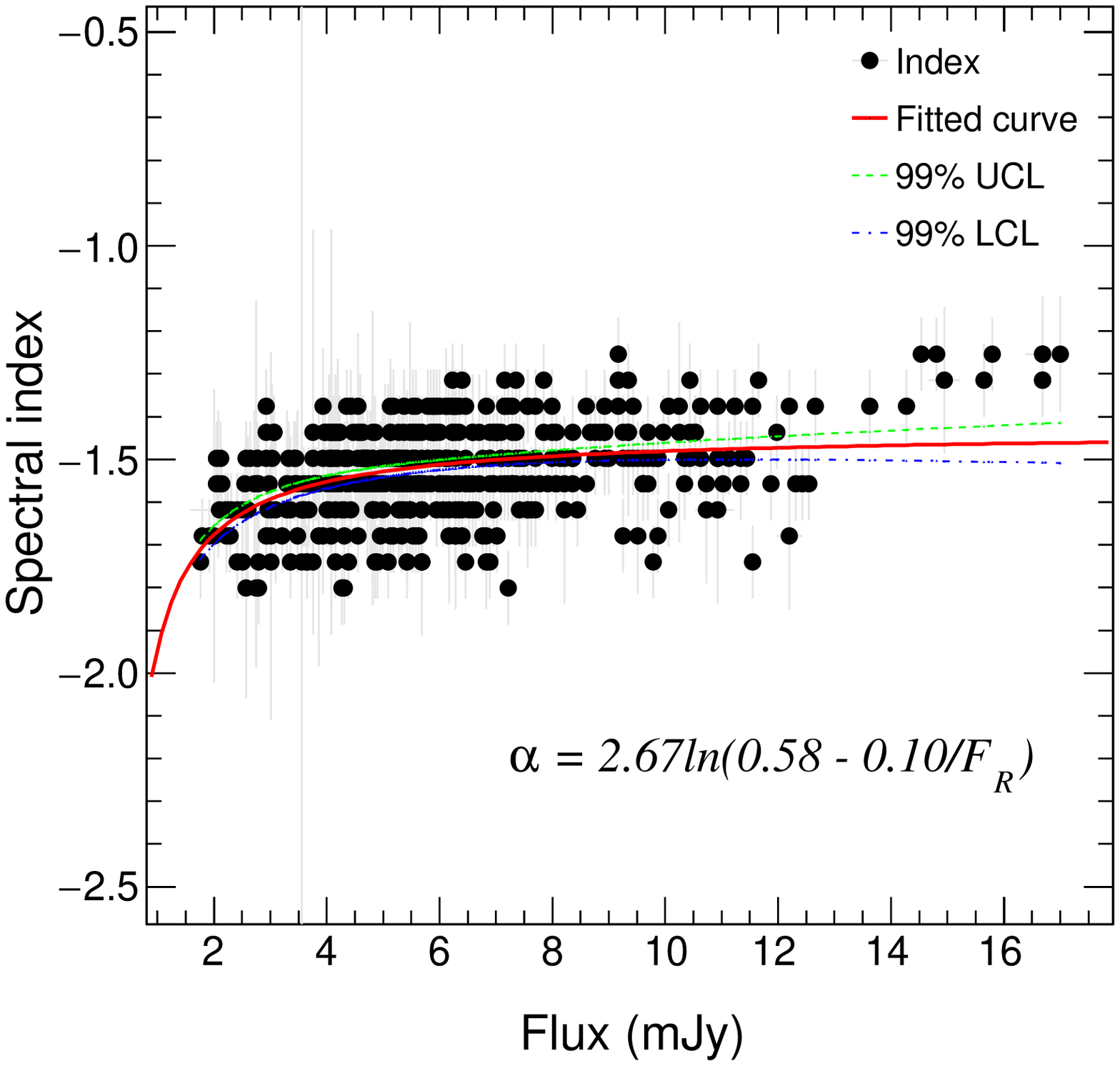}
    \includegraphics[width=0.68\columnwidth]{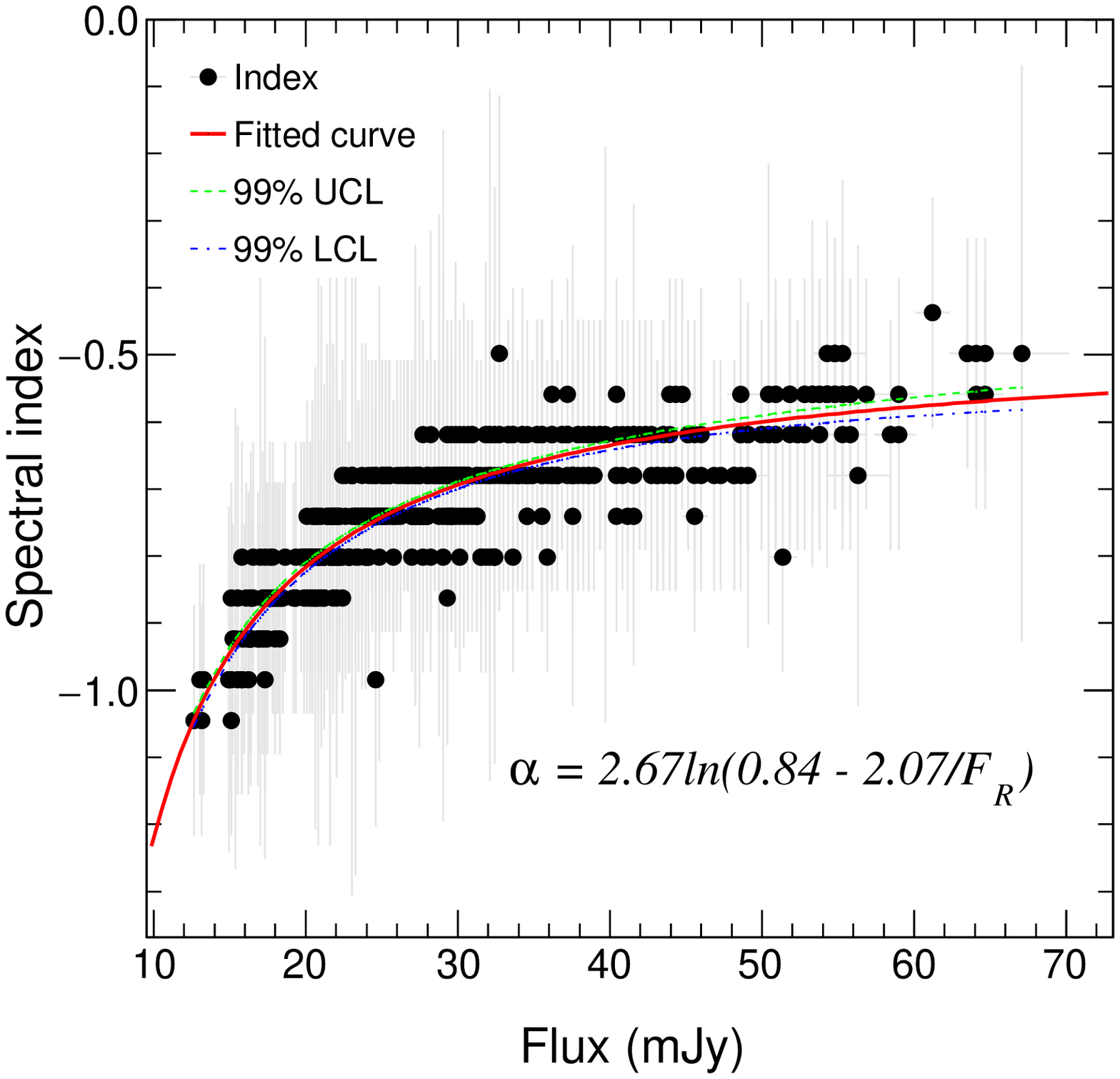}
    \includegraphics[width=0.68\columnwidth]{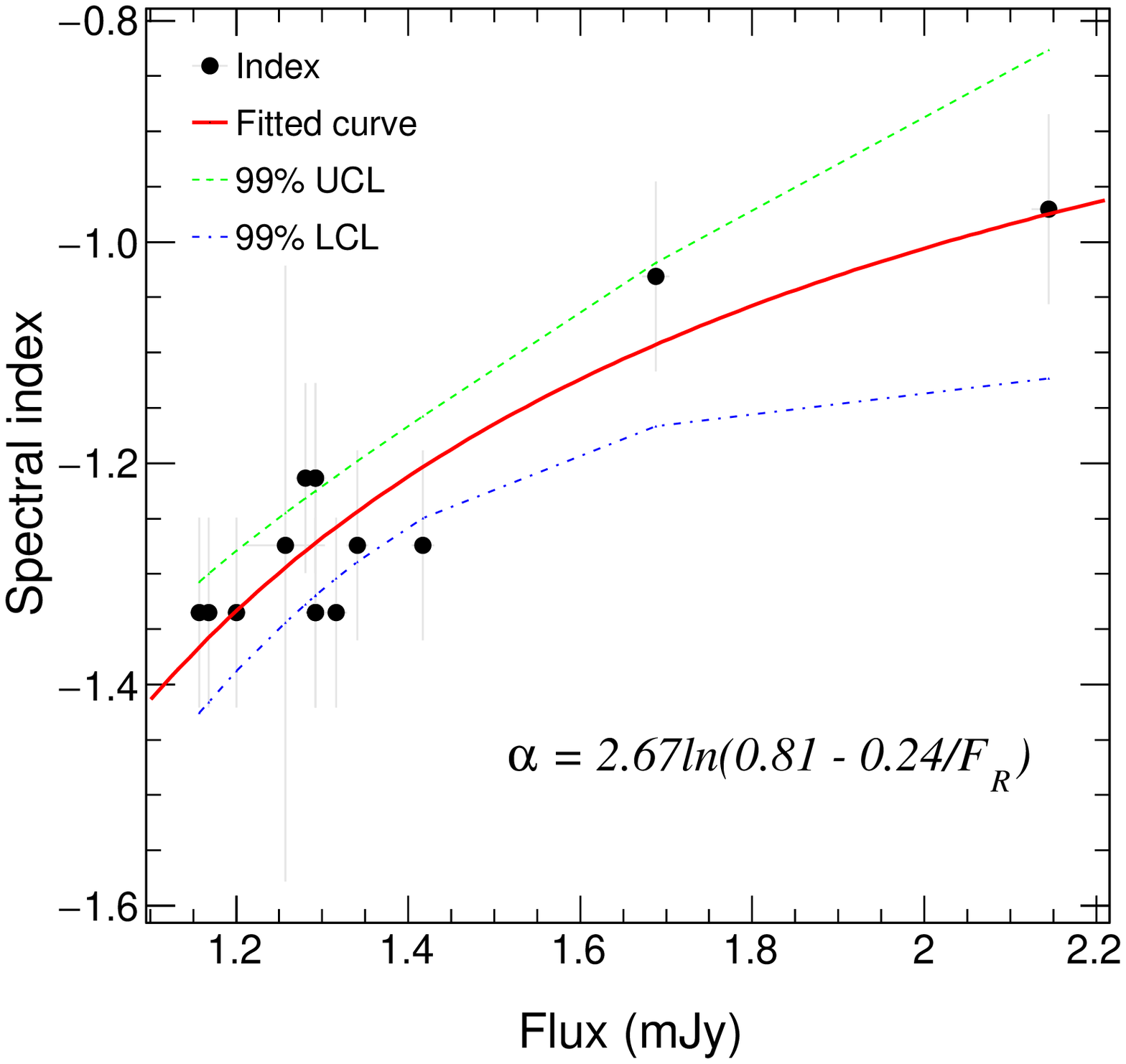}
    \caption{Same as Fig.~\ref{fig:spec-example}, but for BL Lac objects (a) OJ 287, (b) Mrk 421 and (c) H 1219+305 from left to right. }
    \label{fig:spec-bllac2}
\end{figure*}

\begin{figure*}
    \includegraphics[width=0.68\columnwidth]{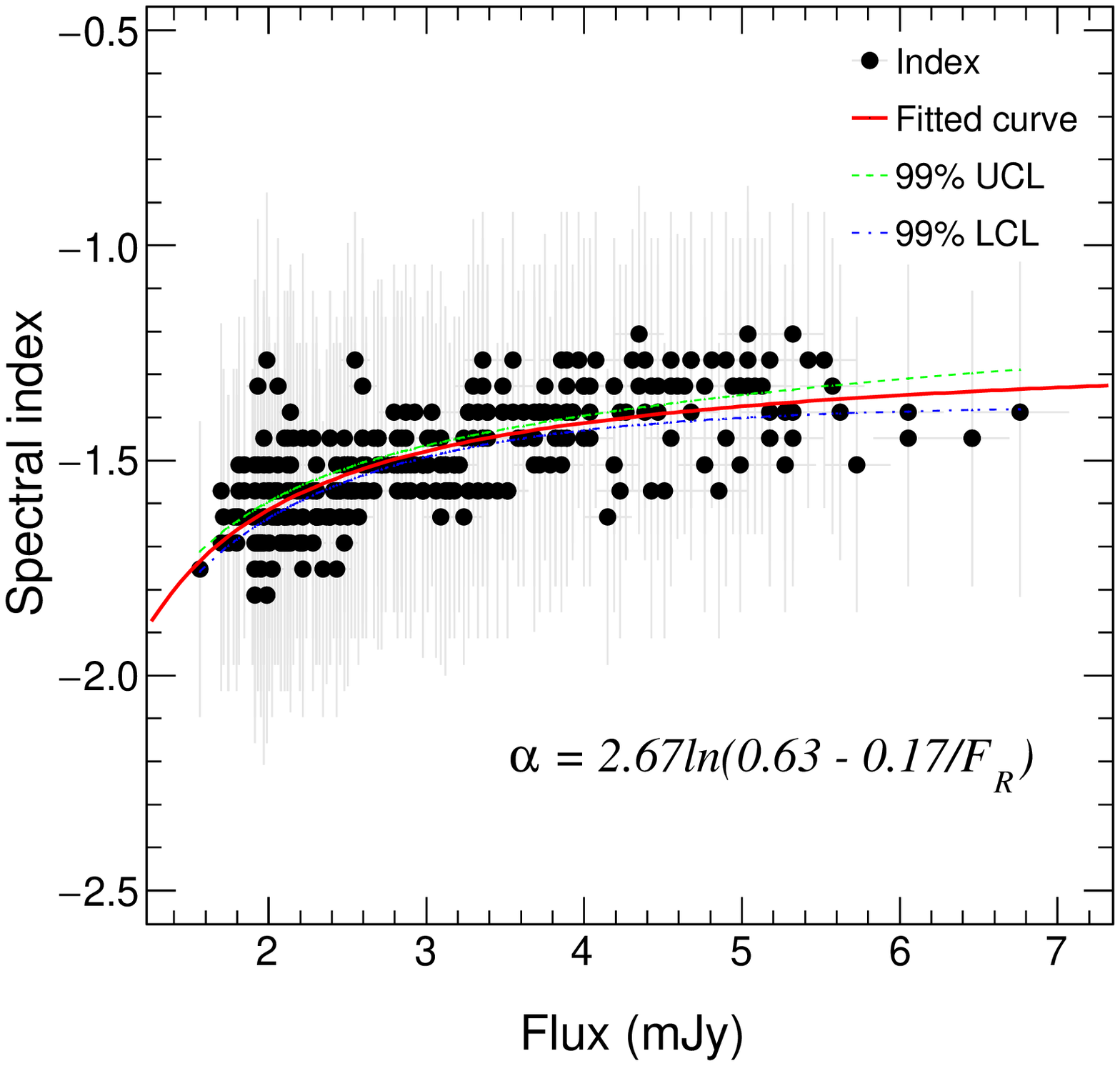}
    \includegraphics[width=0.68\columnwidth]{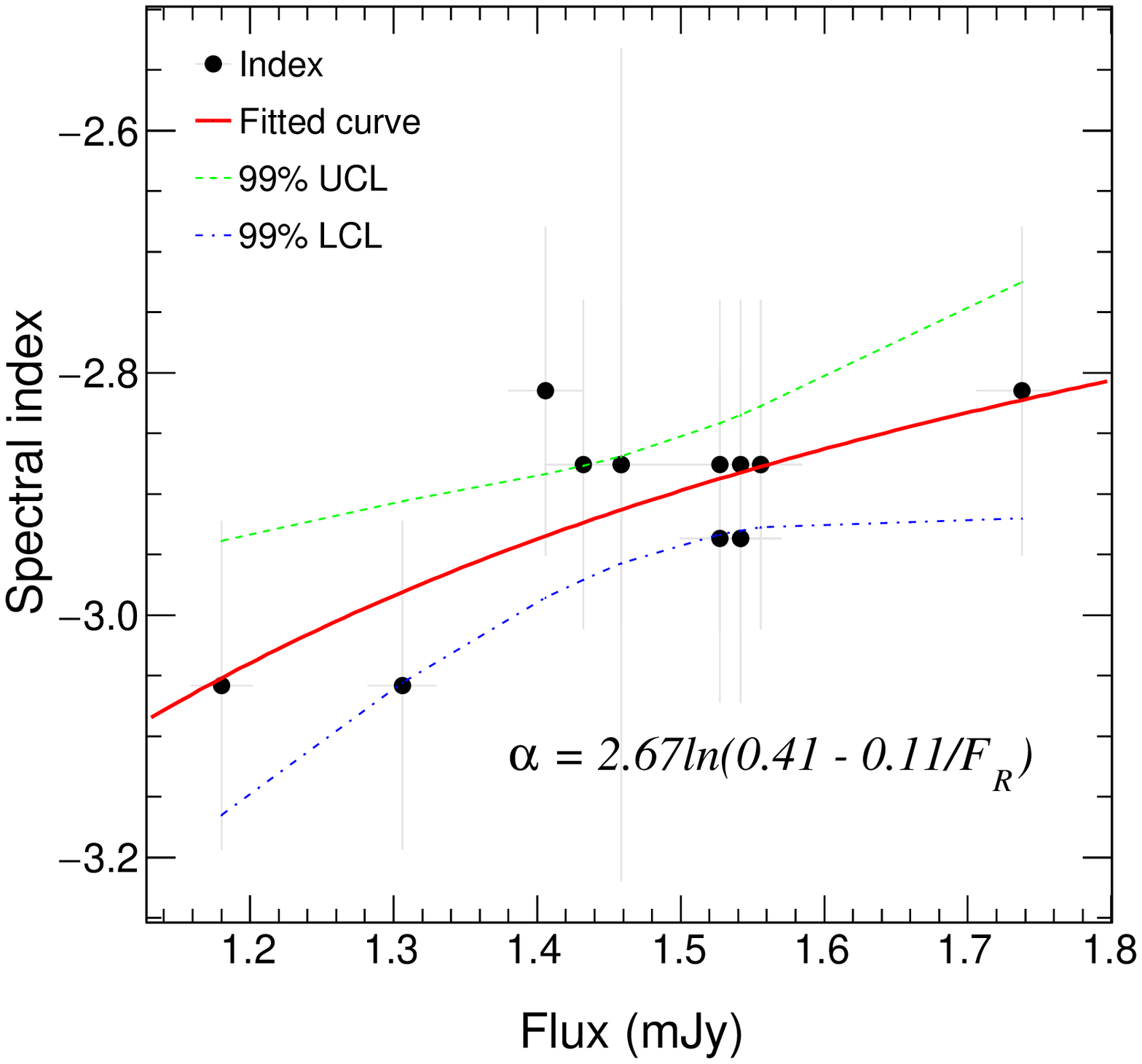}
    \includegraphics[width=0.68\columnwidth]{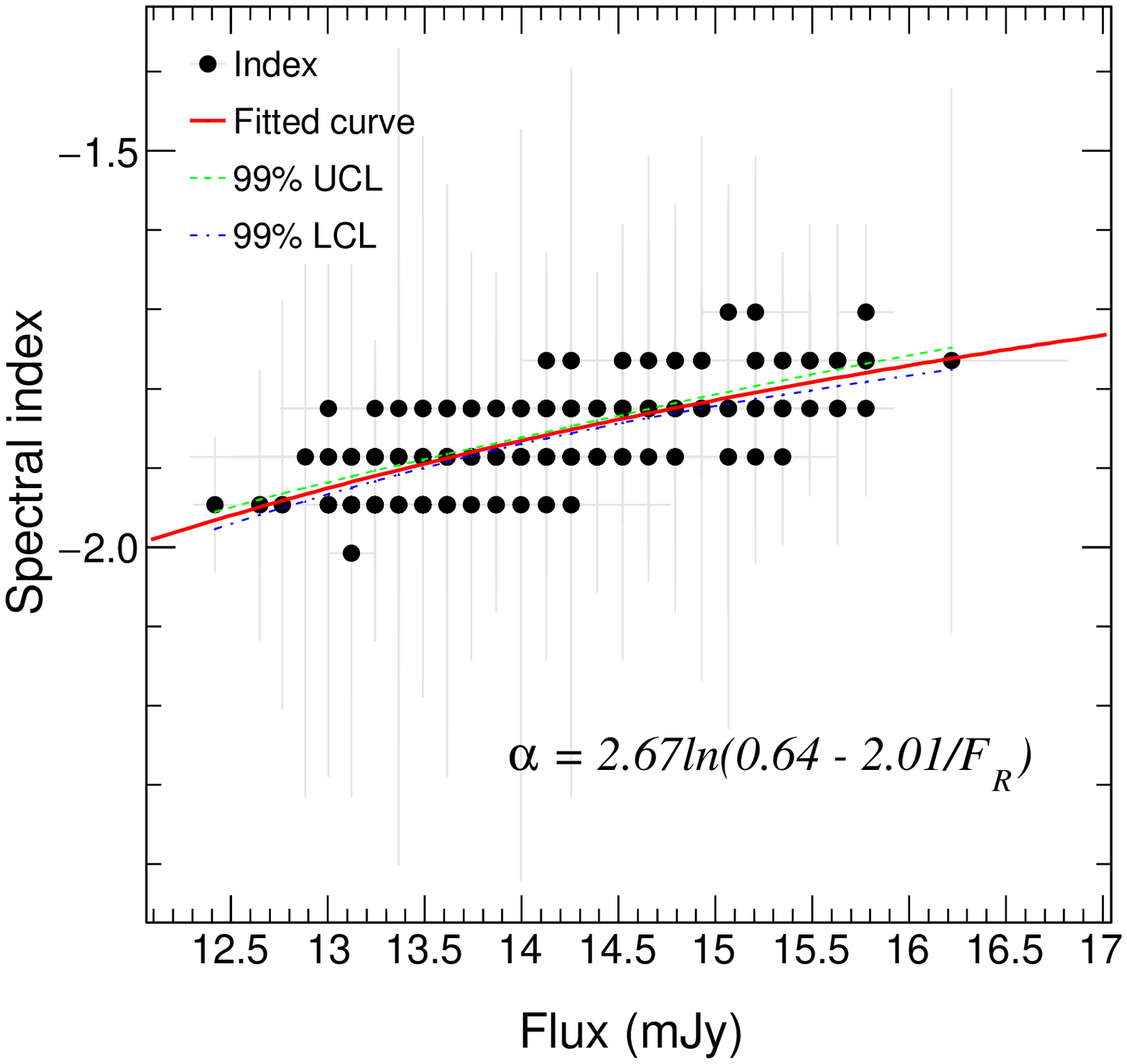}
    \caption{Same as Fig.~\ref{fig:spec-example}, but for BL Lac objects (a) W Com, (b) H 1426+428 and (c) Mrk 501 from left to right.}
    \label{fig:spec-bllac3}
\end{figure*}

\begin{figure*}
    \includegraphics[width=0.68\columnwidth]{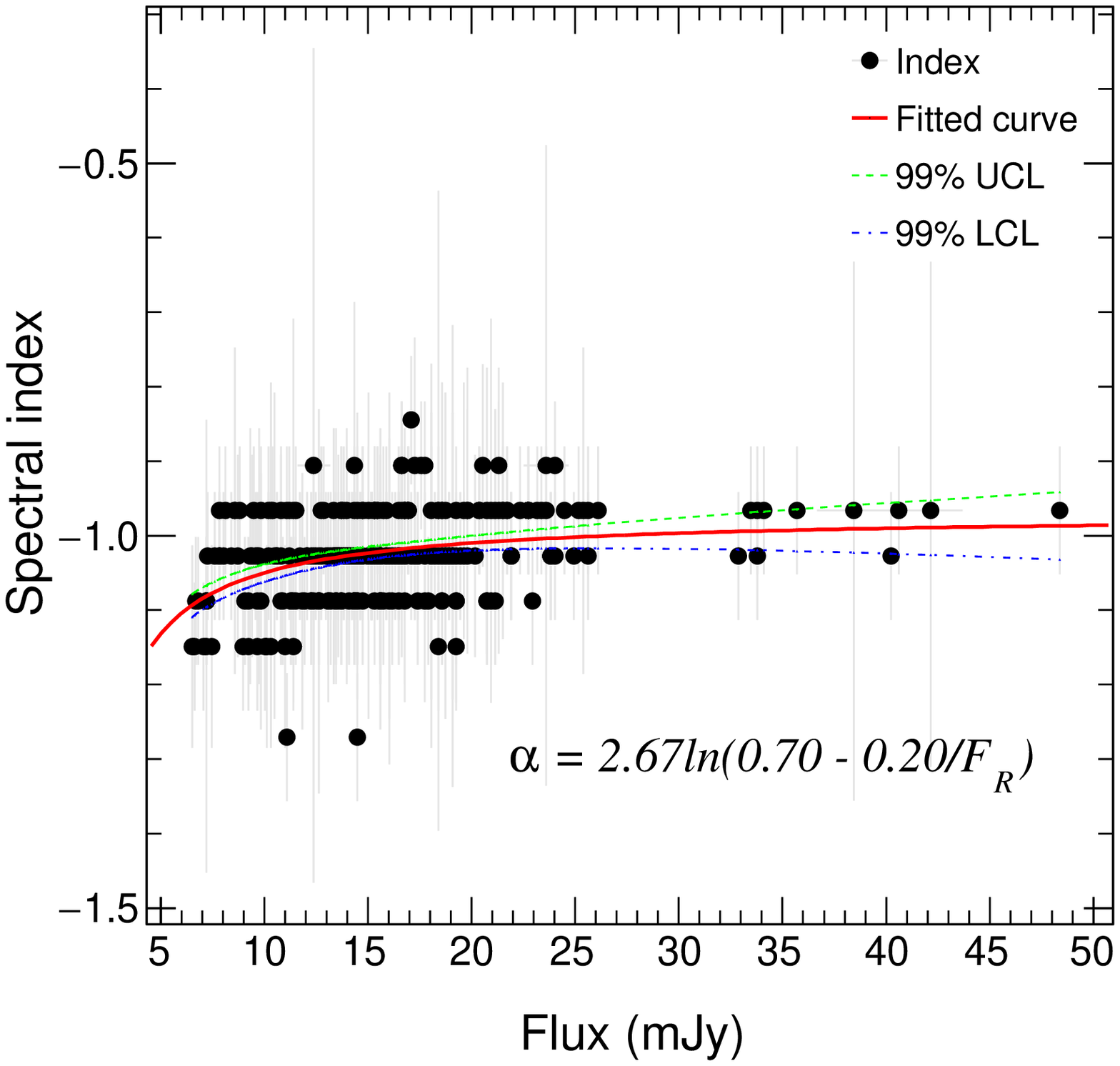}
    \includegraphics[width=0.68\columnwidth]{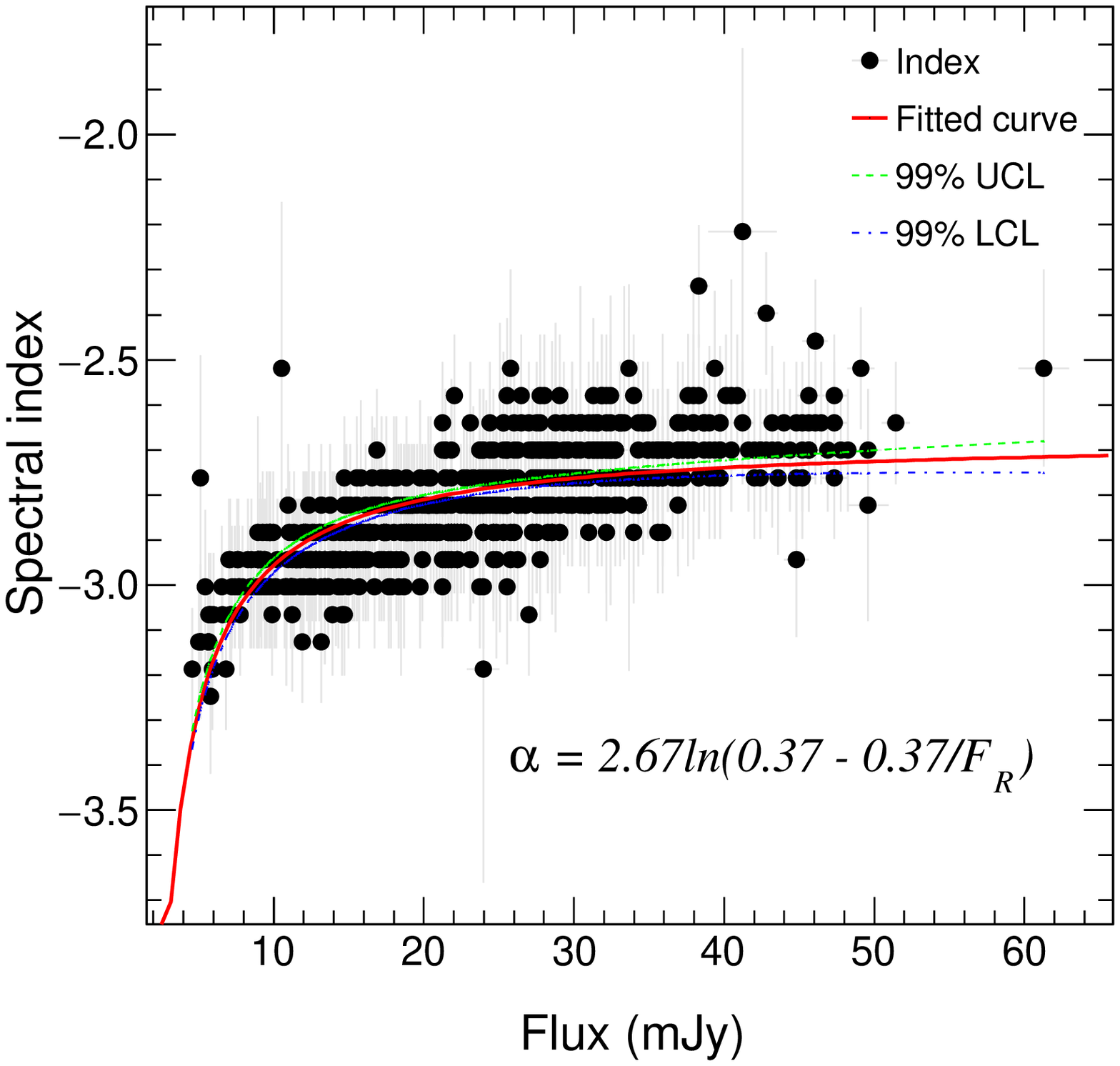}
    \includegraphics[width=0.68\columnwidth]{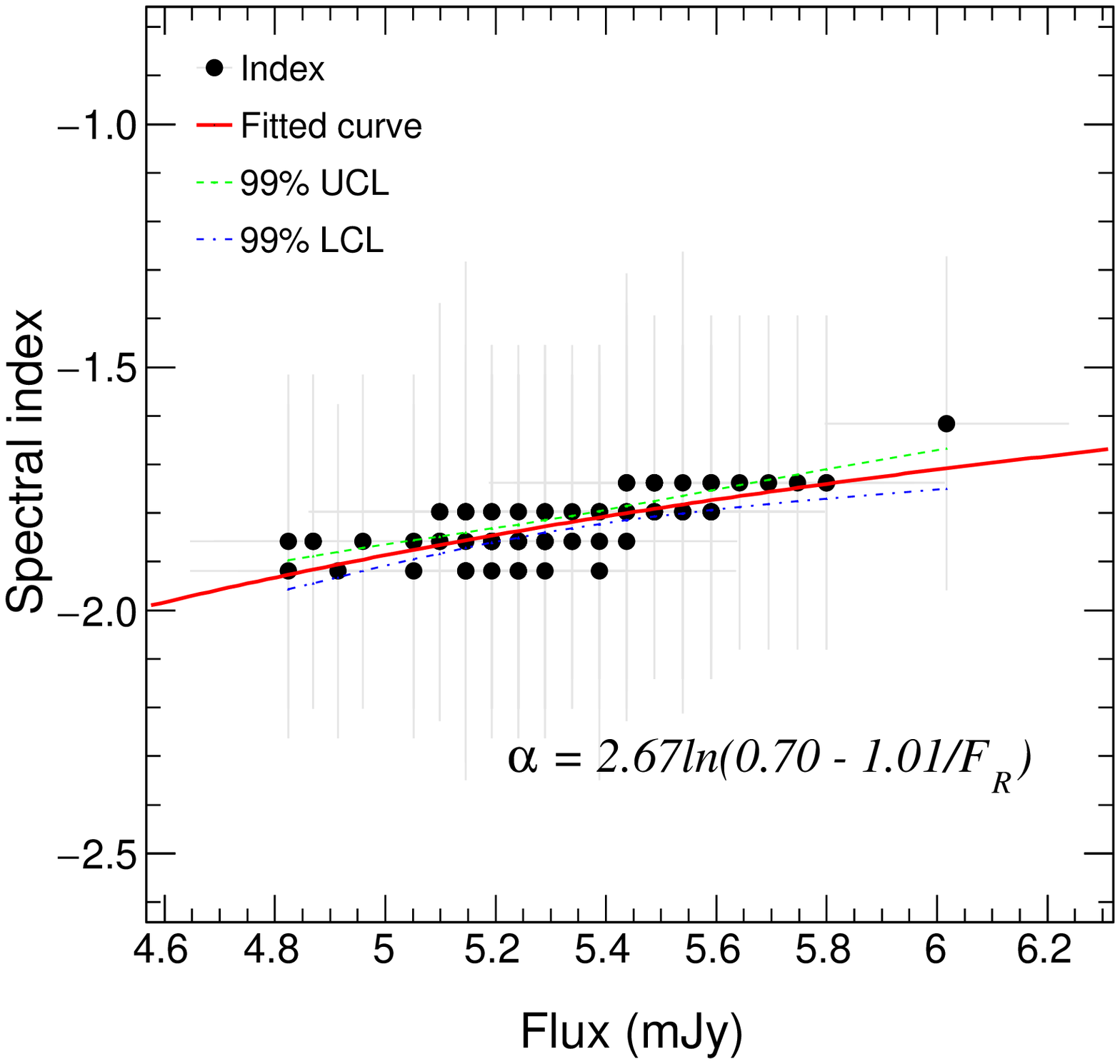}
    \caption{Same as Fig.~\ref{fig:spec-example}, but for BL Lac objects (a) PKS 2155-304, (b) BL Lac and (c) 1ES 2344+514 from left to right.}
    \label{fig:spec-bllac4}
\end{figure*}

\begin{figure*}
    \includegraphics[width=0.68\columnwidth]{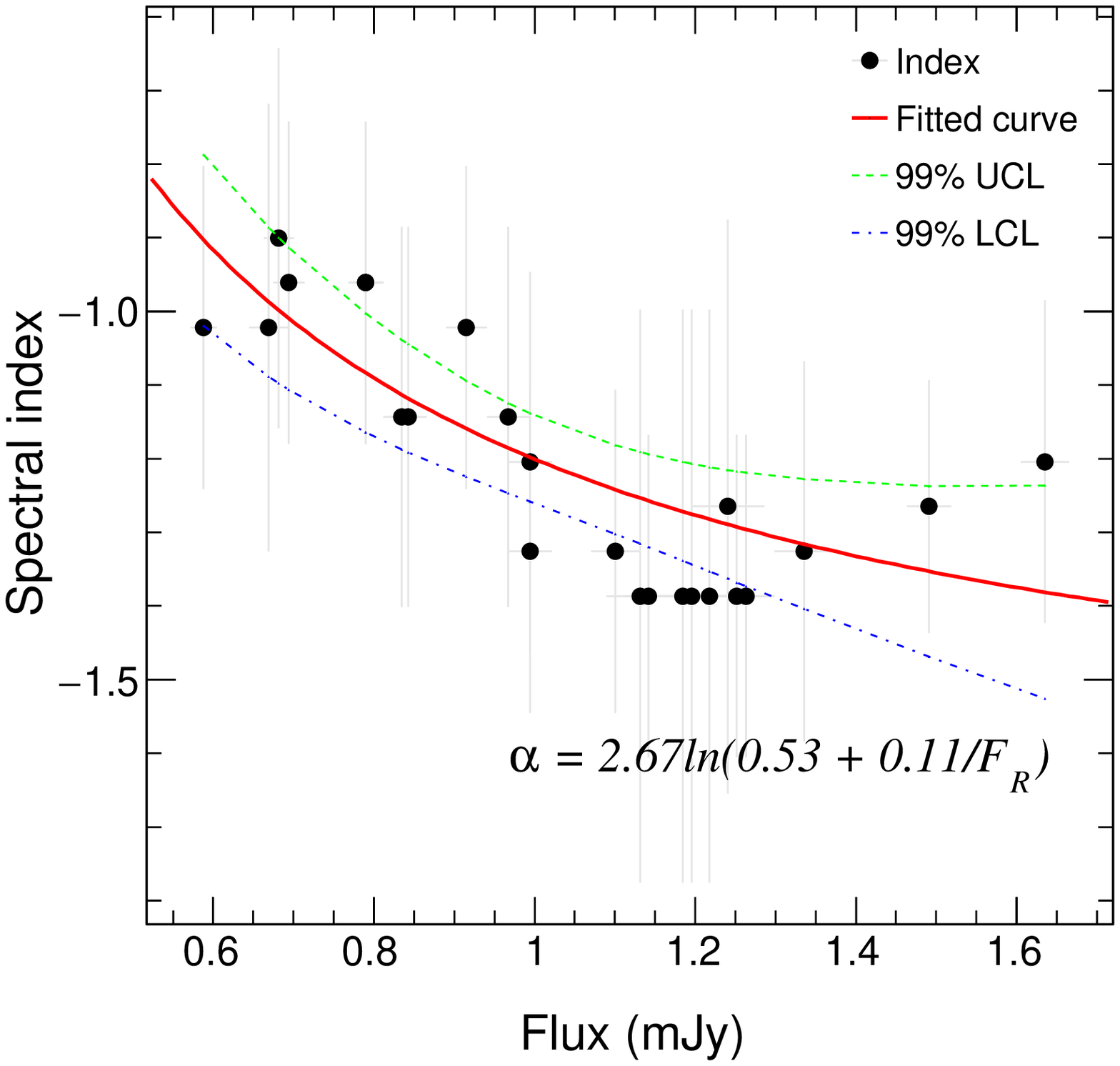}
    \includegraphics[width=0.68\columnwidth]{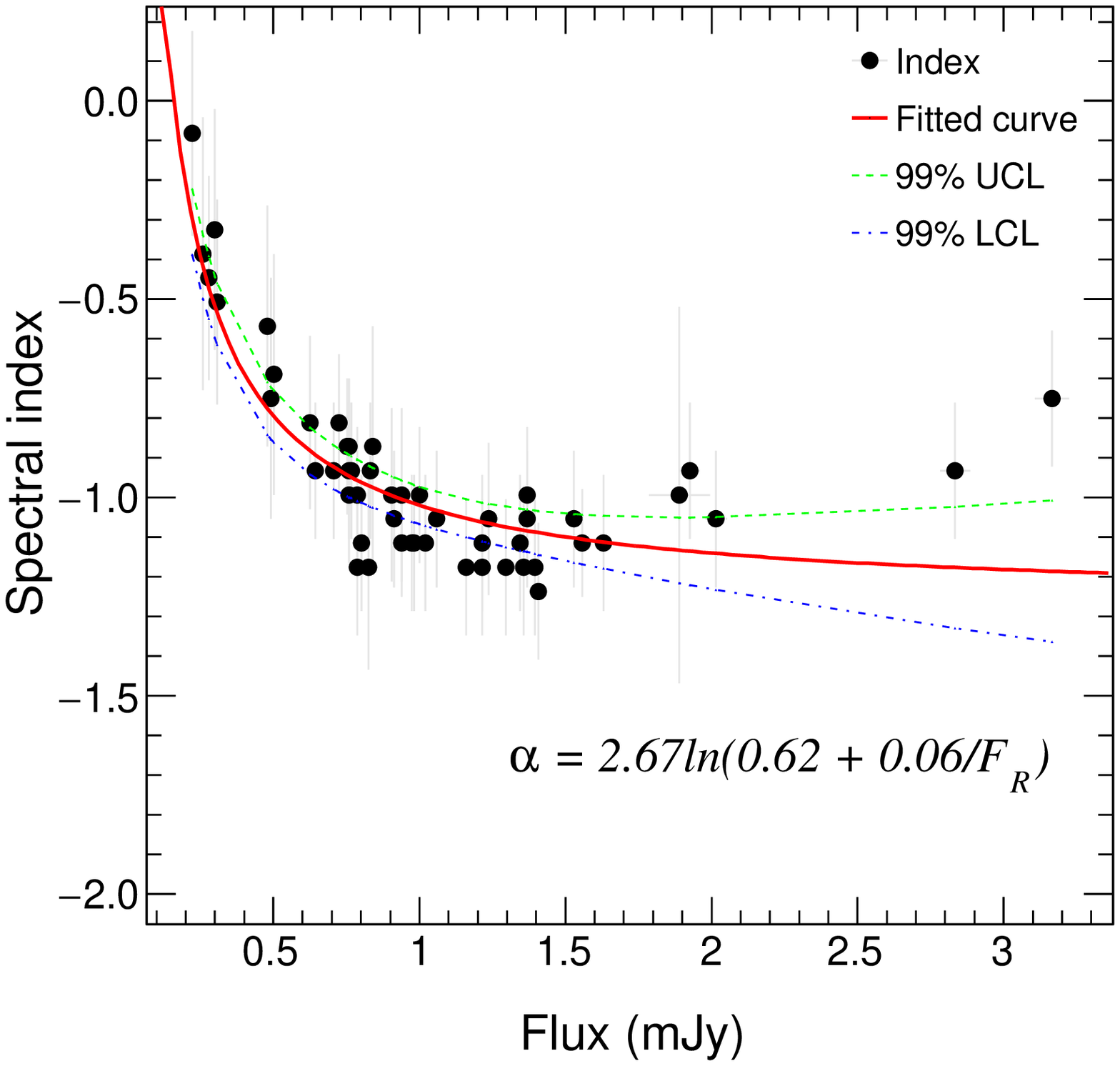}
    \includegraphics[width=0.68\columnwidth]{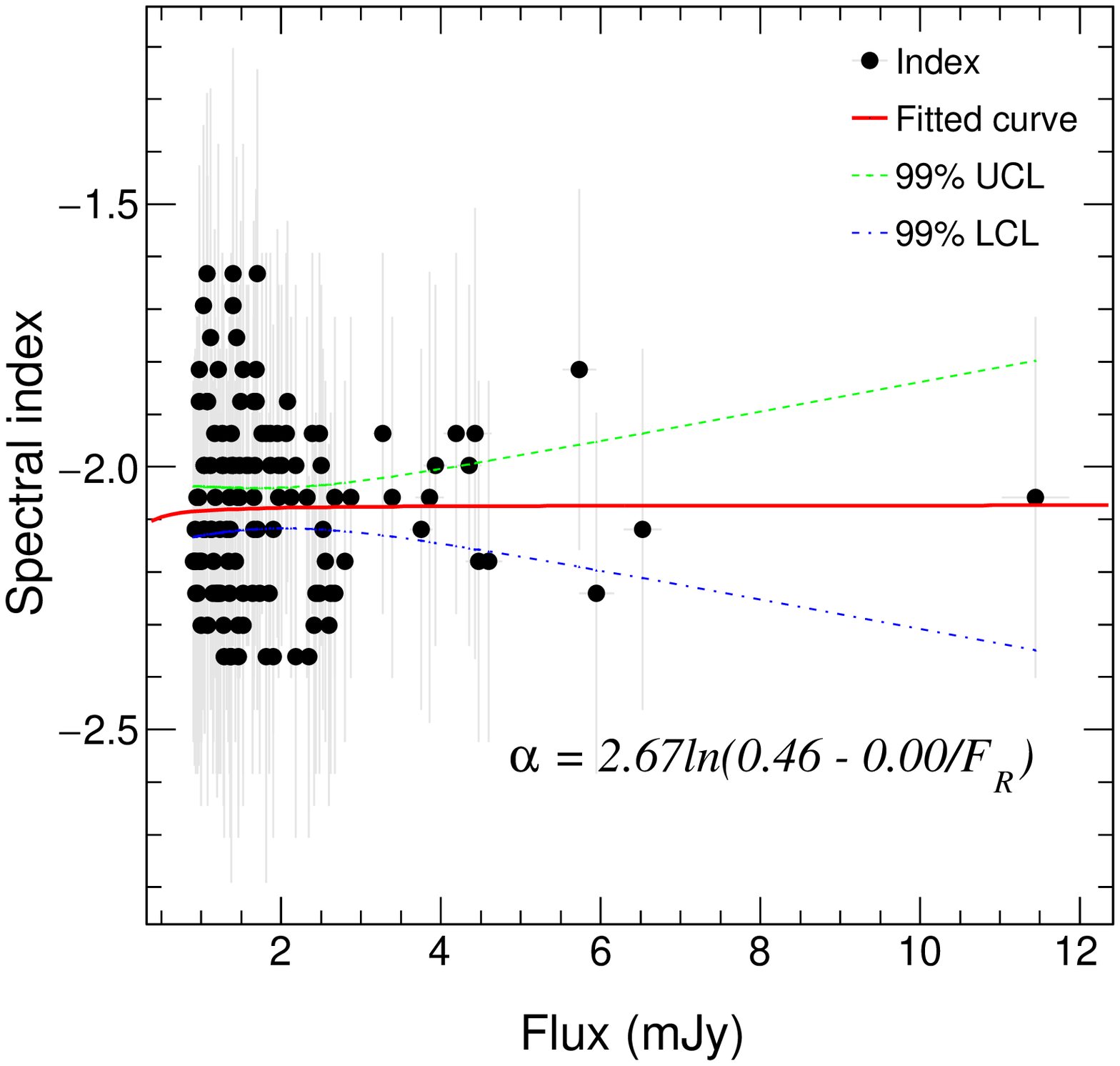}
    \caption{Same as Fig.~\ref{fig:spec-example}, but for FRSQs (a) CTA 26, (b) PKS 0420-014 and (c) PKS 0736+01 from left to right. }
    \label{fig:spec-fsrq1}
\end{figure*}

\begin{figure*}
    \includegraphics[width=0.68\columnwidth]{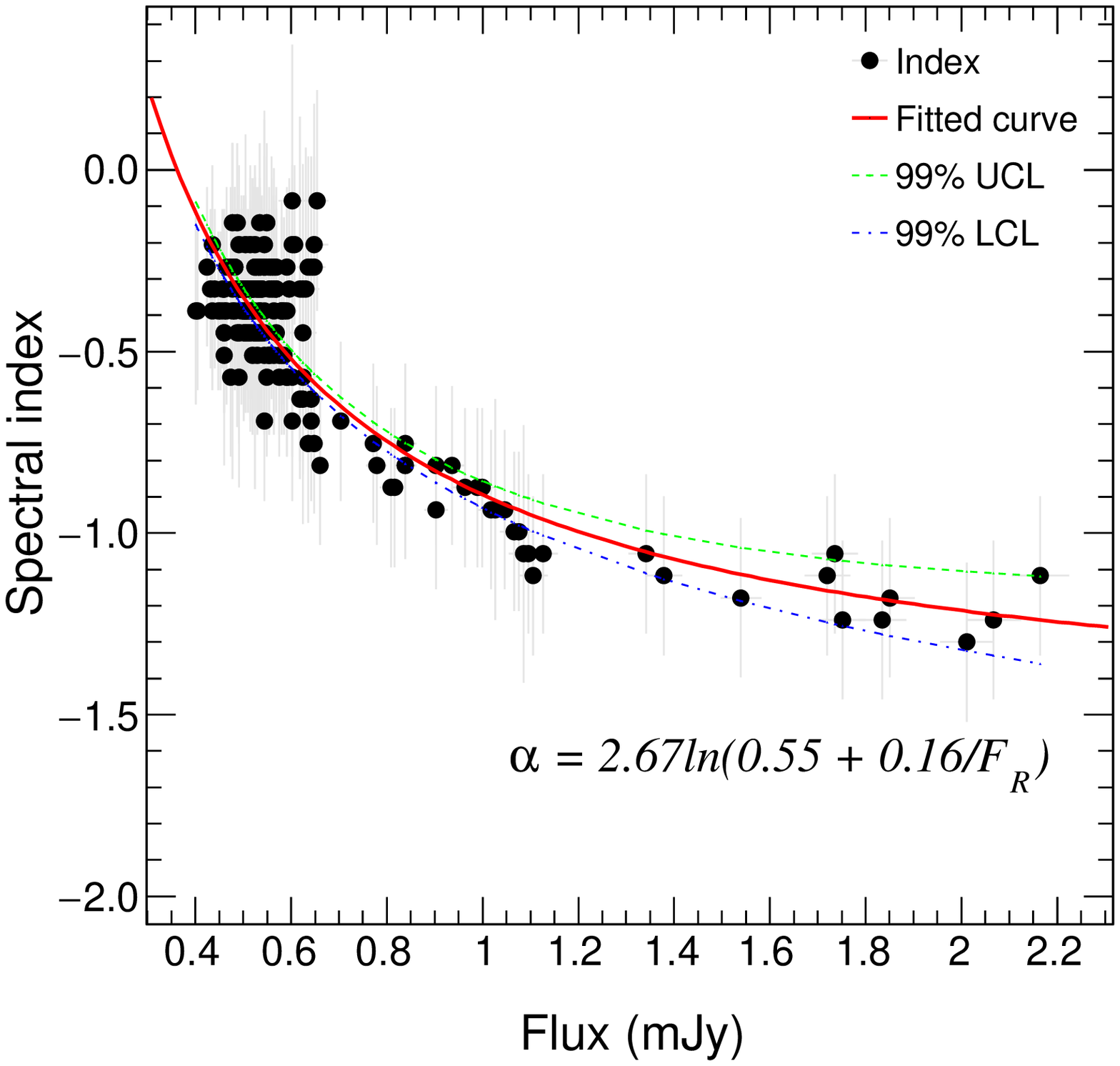}
    \includegraphics[width=0.68\columnwidth]{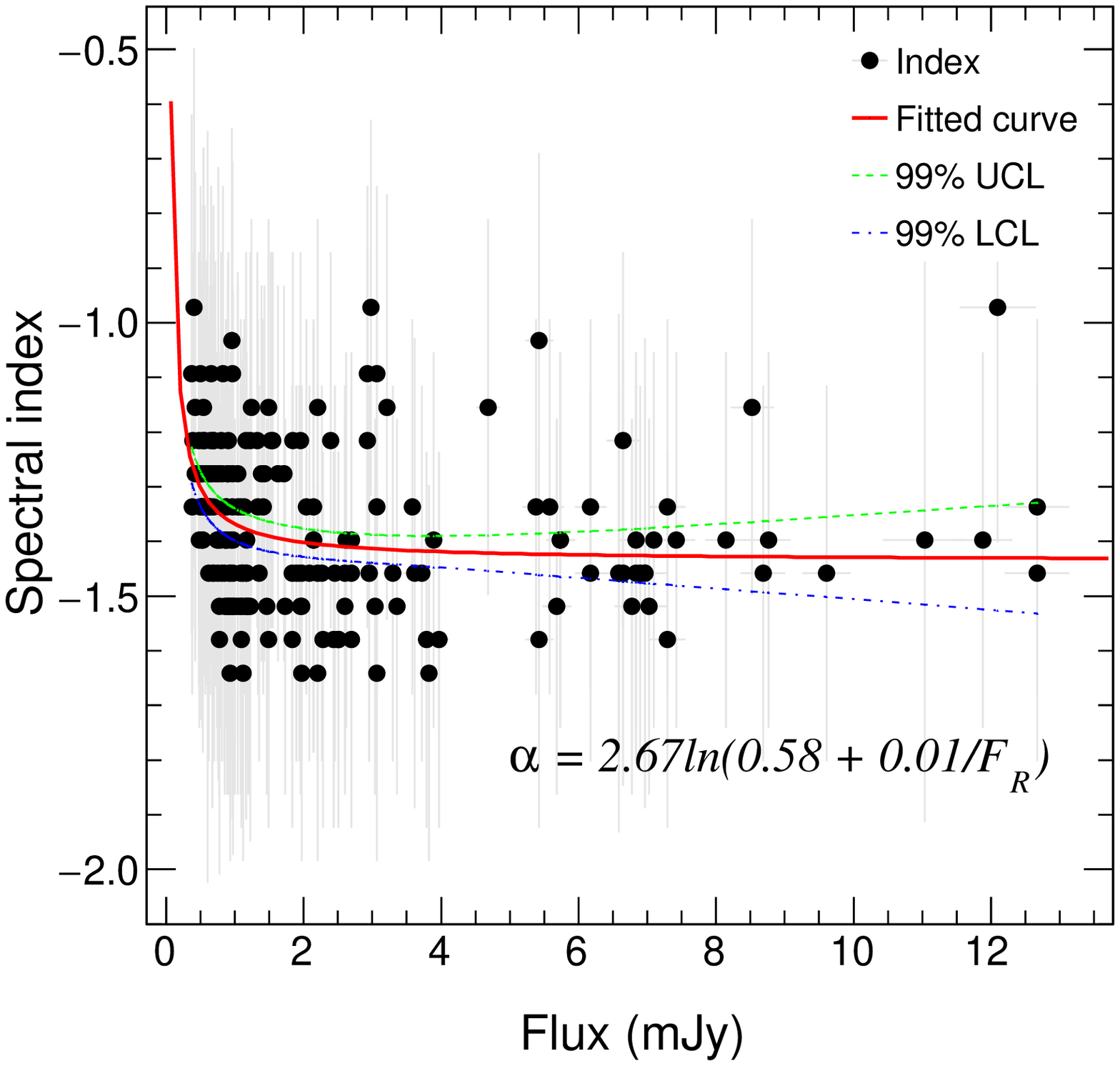}
    \includegraphics[width=0.68\columnwidth]{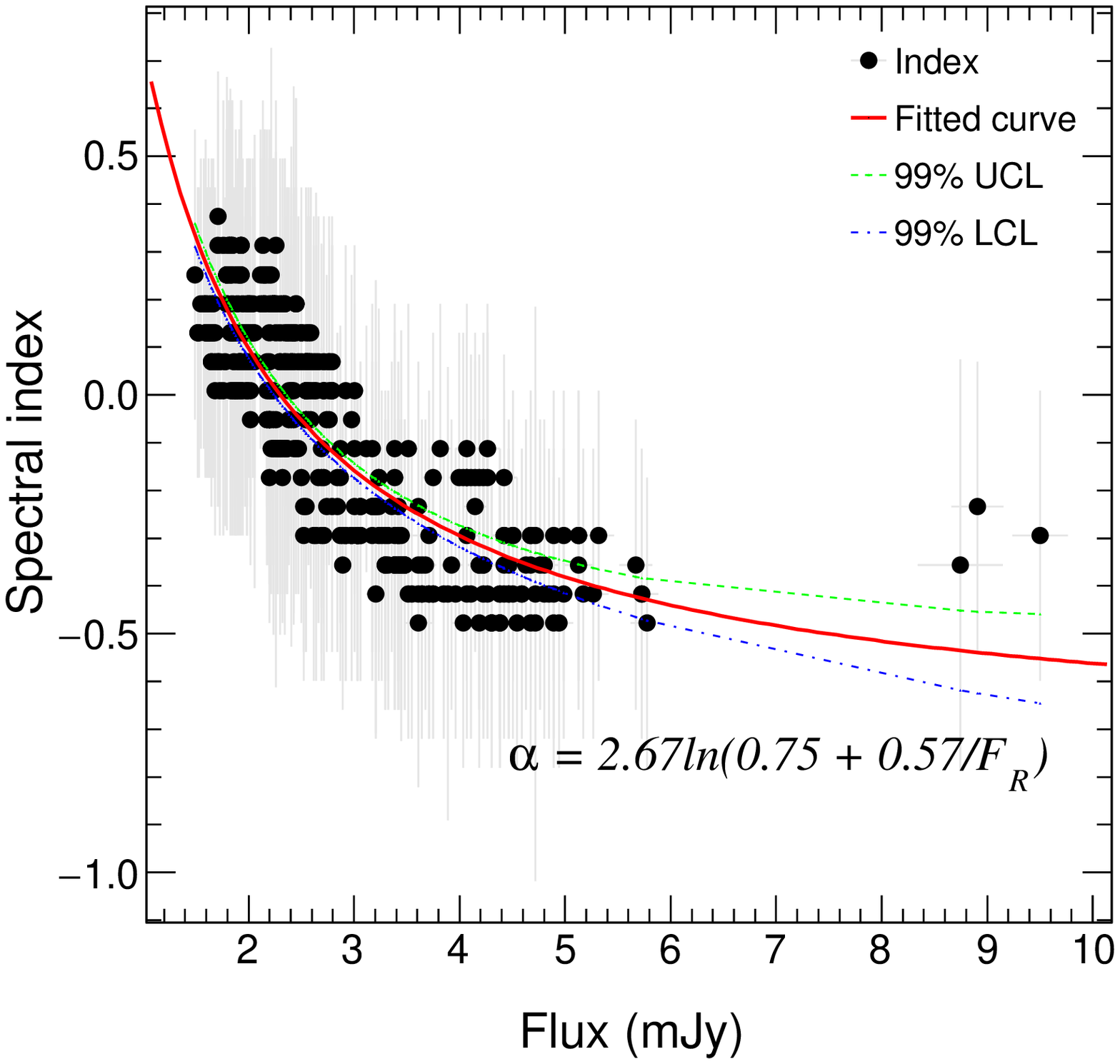}
    \caption{Same as Fig.~\ref{fig:spec-example}, but for FRSQs (a) OJ 248, (b) Ton 599 and (c) PKS 1222+216 from left to right.}
    \label{fig:spec-fsrq2}
\end{figure*}

\begin{figure*}
    \includegraphics[width=0.68\columnwidth]{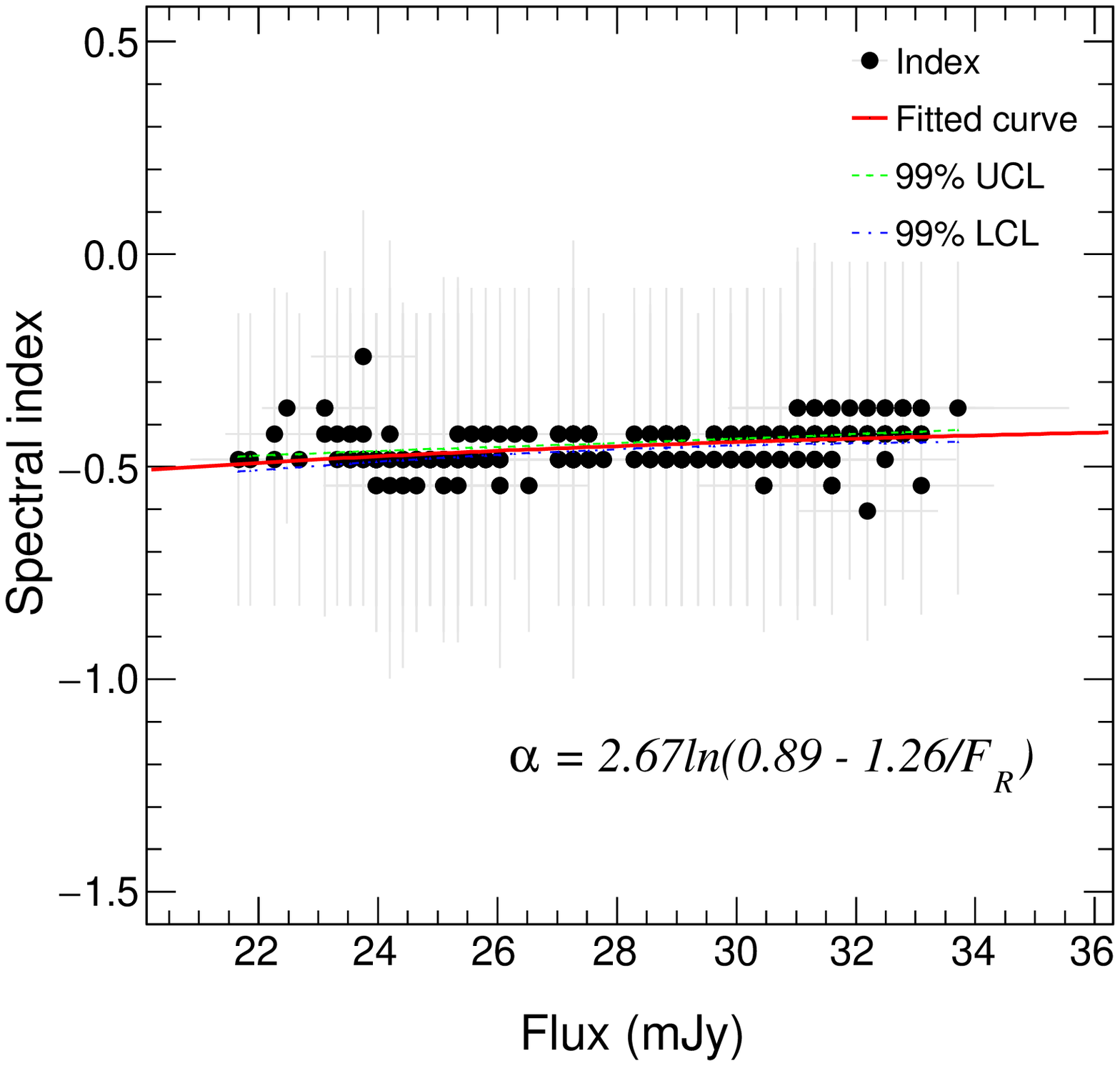}
    \includegraphics[width=0.68\columnwidth]{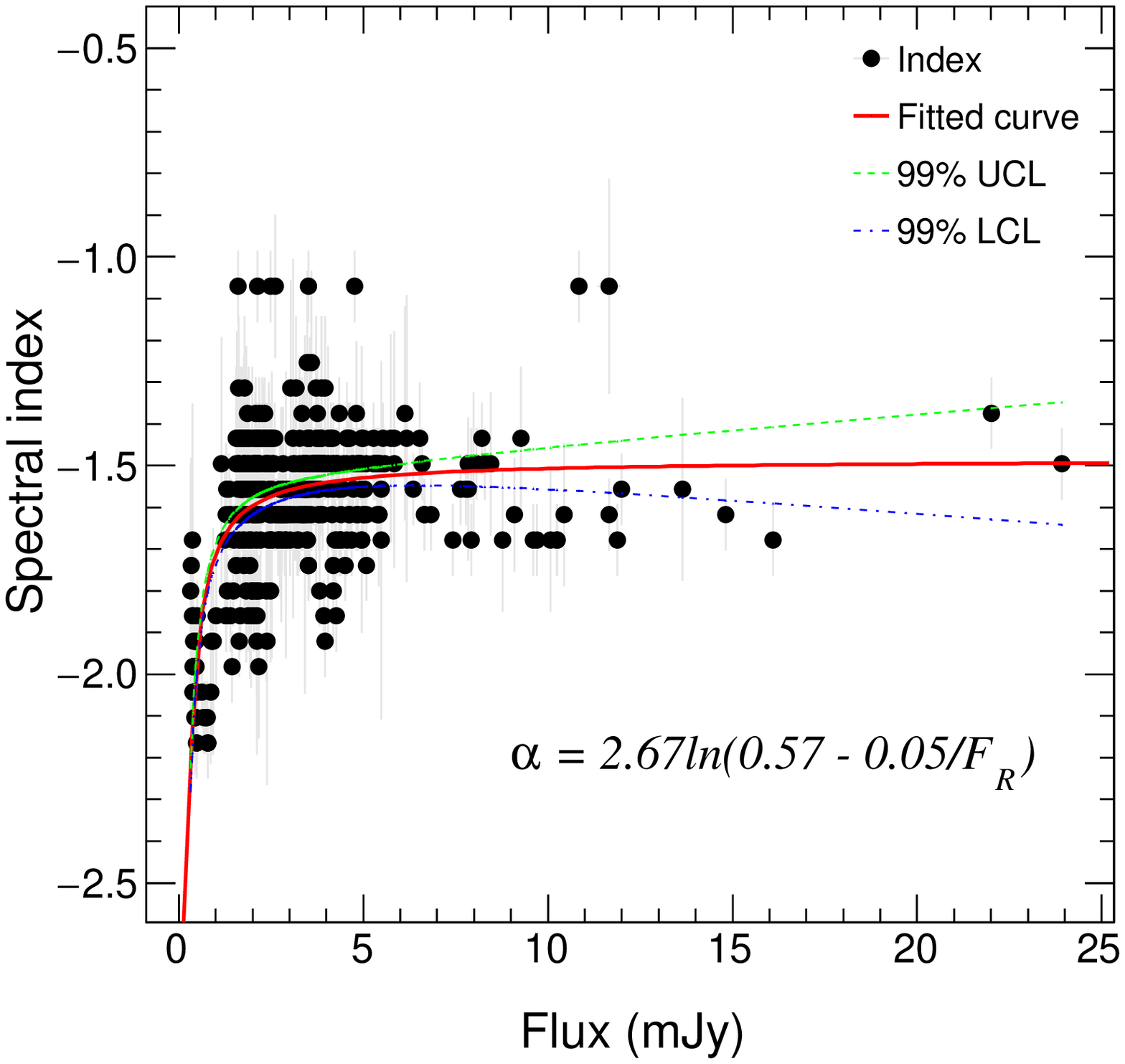}
    \includegraphics[width=0.68\columnwidth]{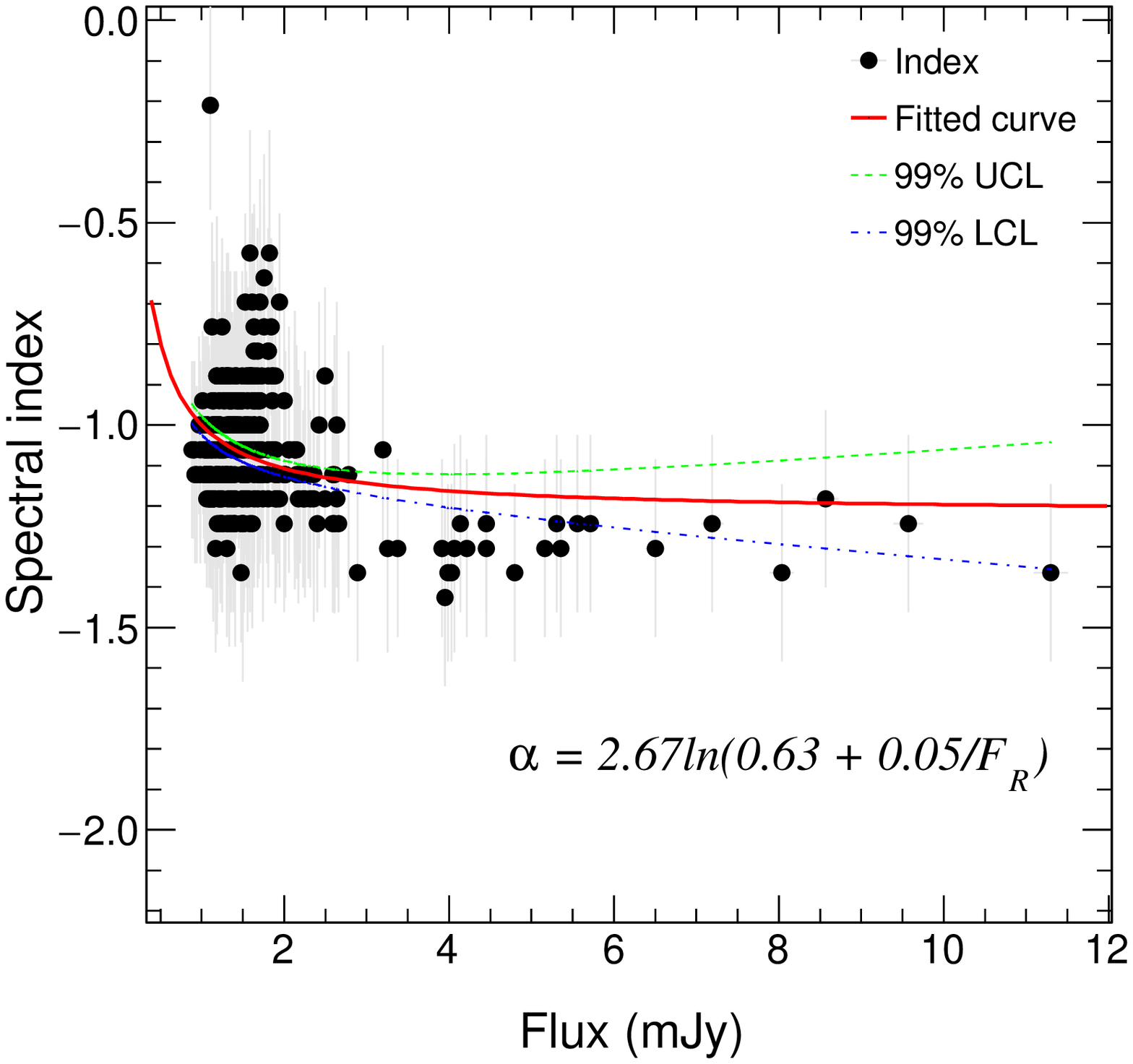}
    \caption{Same as Fig.~\ref{fig:spec-example}, but for FRSQs (a) 3C 273, (b) 3C 279 and (c) PKS 1510-08 from left to right. }
    \label{fig:spec-fsrq3}
\end{figure*}

\begin{figure*}
    \includegraphics[width=0.68\columnwidth]{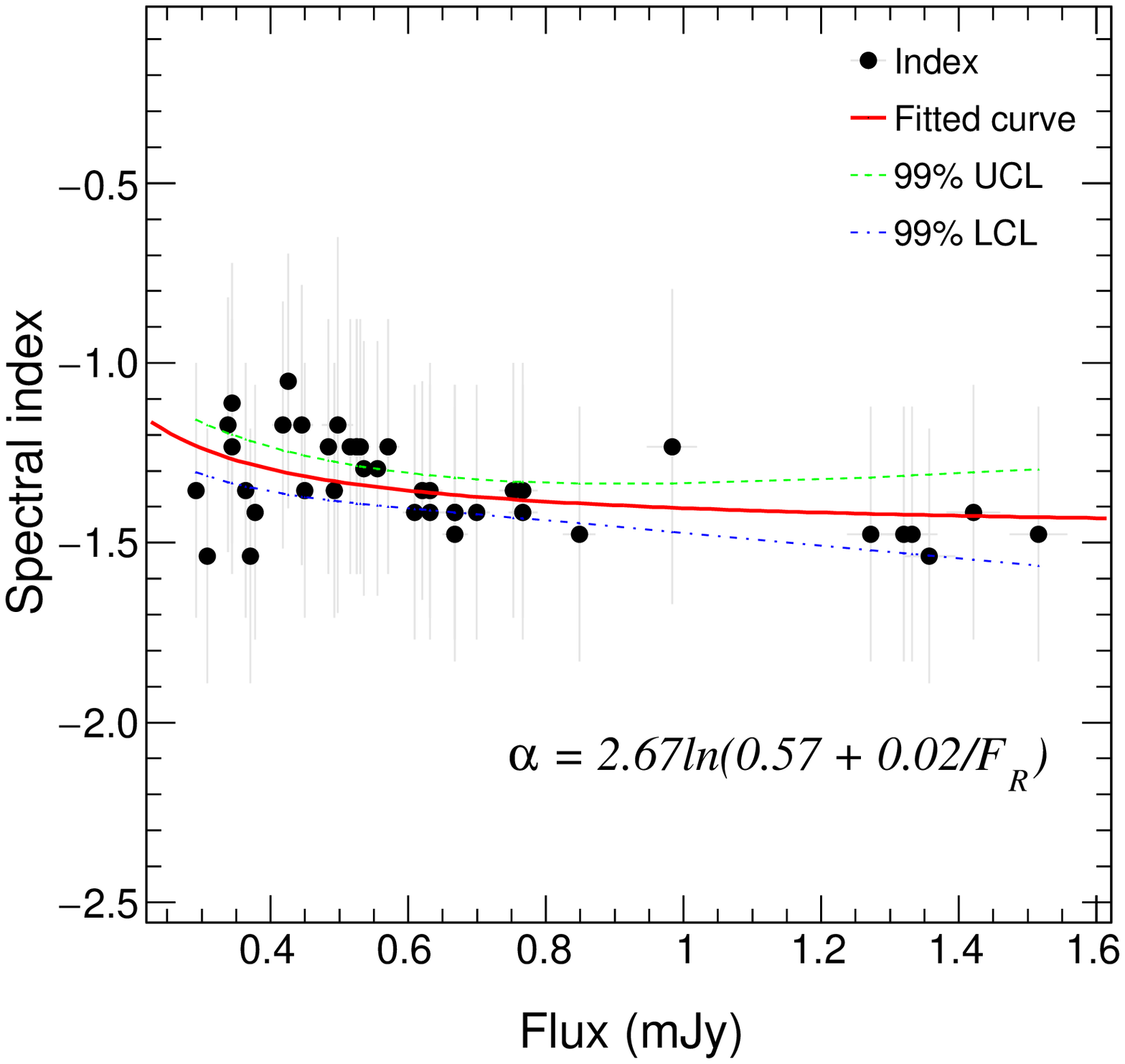}
    \includegraphics[width=0.68\columnwidth]{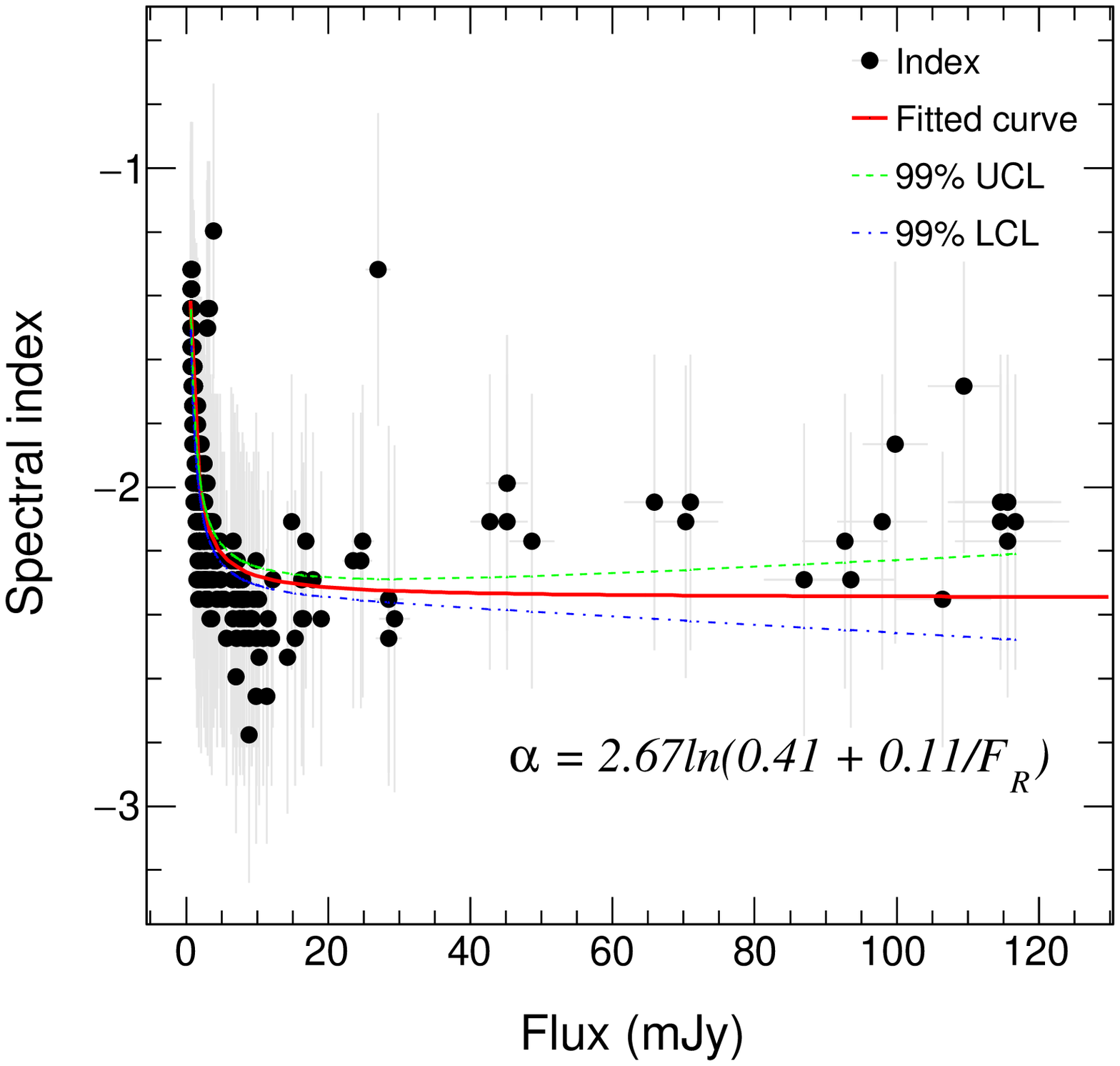}
    \includegraphics[width=0.68\columnwidth]{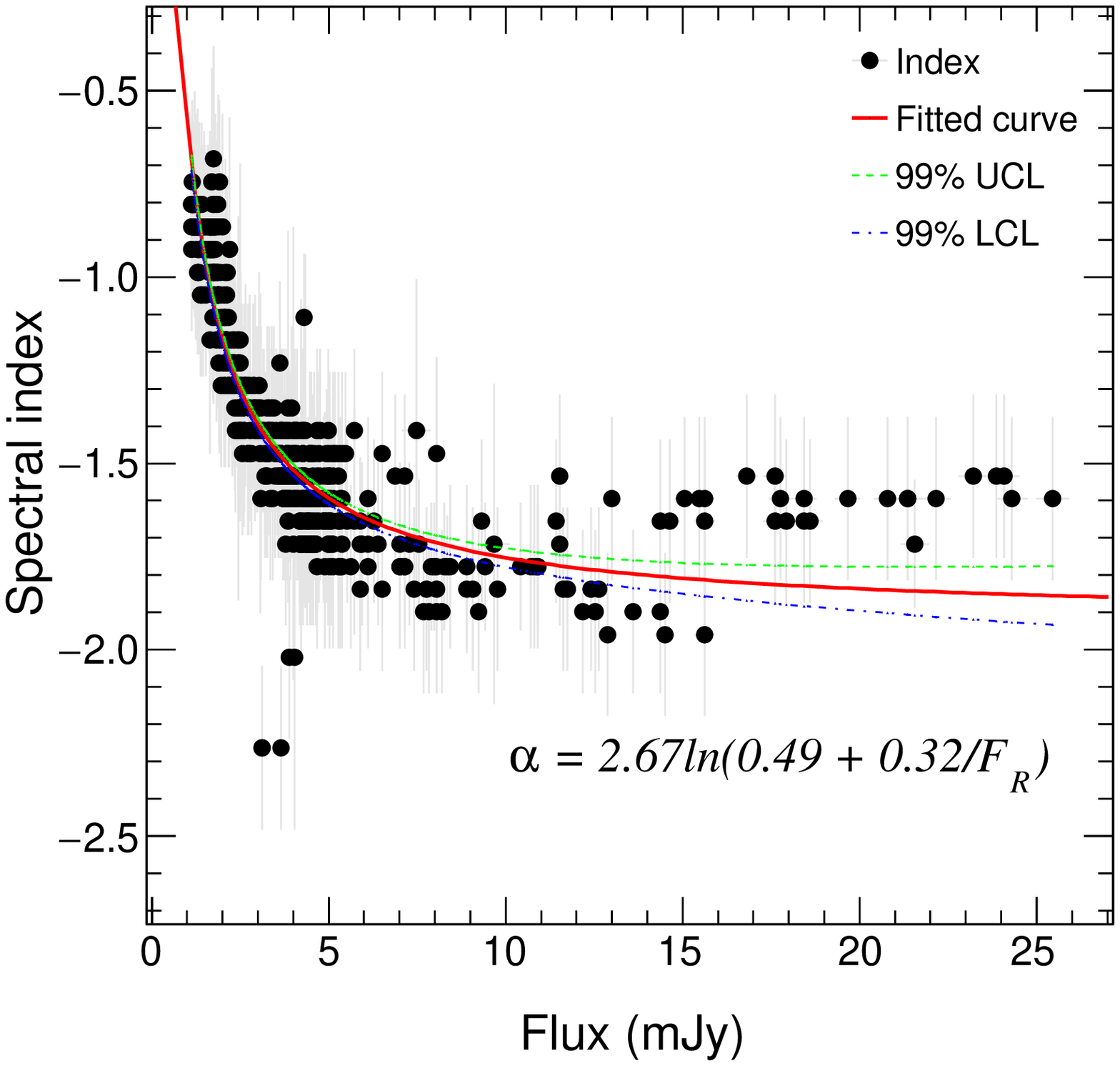}
    \caption{Same as Fig.~\ref{fig:spec-example}, but for FRSQs (a) 3C 345, (b) CTA 102 and (c) 3C 454.3 from left to right. }
    \label{fig:spec-fsrq4}
\end{figure*}

The optical spectral features of other objects are all investigated, and the behaviors are listed in Column 5 of Table~\ref{tab:spec}, and also displayed in Figs.~\ref{fig:spec-bllac0} -~\ref{fig:spec-bllac4} for BL Lac objects and Figs.~\ref{fig:spec-fsrq1} -~\ref{fig:spec-fsrq4} for FSRQs. One can notice that the variation trends of spectral index $\alpha$ with flux $F_{R}$ exhibit BSWB or RSWB trends, and do not seem to follow linear relations. Therefore, we fit the spectral behavior by means of equations~(\ref{Equ:fitln}) and~(\ref{Equ:fitpol1}) for each source, and only superimpose the fitting results with equation~(\ref{Equ:fitln}) on the corresponding figures.
The $\chi^{2}$ values fitted by equation~(\ref{Equ:fitln}) for 20 sources are lower or equal than those by equation~(\ref{Equ:fitpol1}).
For 24 sources, equation~(\ref{Equ:fitln}) has lower or equal BICs than equation~(\ref{Equ:fitpol1}).
Only 3 sources has higher BICs with equation~(\ref{Equ:fitln}) than equation~(\ref{Equ:fitpol1}).
White's test for equation~(\ref{Equ:fitln}) show that there are 19 sources with $p$-value $>$ 0.01, which means that the heteroscedasticity is not statistically significant. In general, the non-linear equation~(\ref{Equ:fitln}) can well describe the optical spectral features, and is superior to the linear equation~(\ref{Equ:fitpol1}).

\section{Discussion}

In total, the features of optical spectral behavior of 27 blazars have been explored, including 14 BL Lacs and 13 FSRQs. Generally, they follow the BSWB or RSWB trends well. Only one exception PKS 0736+01 exhibits, on average, the stable-when-brighter (SWB) trend.
In the past, the BWB or RWB trends were widely used to describe the spectral characteristics of blazars.
However, an increasing number of hints indicate that BWB and RWB trends are insufficient to represent the spectral characteristics of blazars.
It was observed that, as the source brightens and exceeds a certain value, the BWB or RWB trend was subsequently followed by a saturation phenomena \citep[e.g.][]{villata06,zhang15,zhang21,sarkar19,safna20,xiong20,otero22}. That is, the color keeps stable when the source brightens above a certain level (stable-when-brighter, SWB).
Recently, \cite{zhang22} found two new universal optical spectral behaviors, i.e., BSWB and RSWB trends,
and inferred that both of them as well as the SWB trend can be regarded as special cases of BSWB and RSWB trends, respectively.
The results of this work further confirm  the existence of two spectral behaviors in the optical region, namely BSWB and RSWB trends.

The previous results suggest that most FSRQs exhibit a RWB trend, while most BL Lacs objects show BWB behavior \citep[e.g.][and references therein]{safna20}. Also, there exist exceptional cases, for example, all 6 BL Lacs show no BWB trends according to the data observed by the Rapid Eye Mounting (REM) telescope during April 2005 to June 2012 \citep{sandrinelli14}, and another example is that only one out of 29 FSRQs in the Sloan Digital Sky Survey stripe 82 region exhibits RWB trend \citep{gu11}.
\cite{zhang22} analyzed the spectral behaviors of 53 Fermi blazars, and their results showed that the vast majority of FSRQs exhibit RSWB trends, however, BL Lacs show no clear preference between BSWB and RSWB trends in their sample.

When we examine the subclass of blazars in detail in this work, we find that 13 out of the 14 BL Lacs obey the BSWB trend, and only one exception, BL Lac object AO 0235+164, follows RSWB trend (see Figs.~\ref{fig:spec-bllac0} -~\ref{fig:spec-bllac4}). For 13 FSRQs, there are 10 objects following the RSWB trend and 2 (3C 273 and 3C 279) following the BSWB trend (see Figs.~\ref{fig:spec-fsrq1} -~\ref{fig:spec-fsrq4}). This means that BL Lacs are more inclined to exhibit BSWB trend and FSRQs are more inclined to exhibit RSWB trend.
It seems that the spectral behavior of an object is independent of whether it is a TeV source.

\begin{figure}
    \includegraphics[width=1.0\columnwidth]{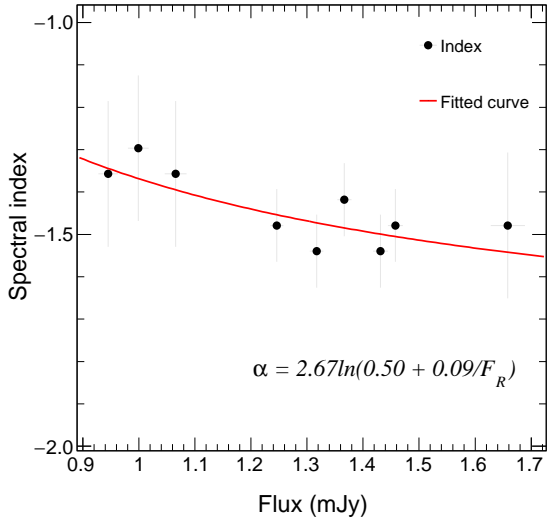}
    \caption{Same as Fig.~\ref{fig:spec-example}, but for FRSQ B2 1633+382 during MJD 58281--58293.}
    \label{fig:spec-shortscale}
\end{figure}

\begin{figure}
    \includegraphics[width=1.0\columnwidth]{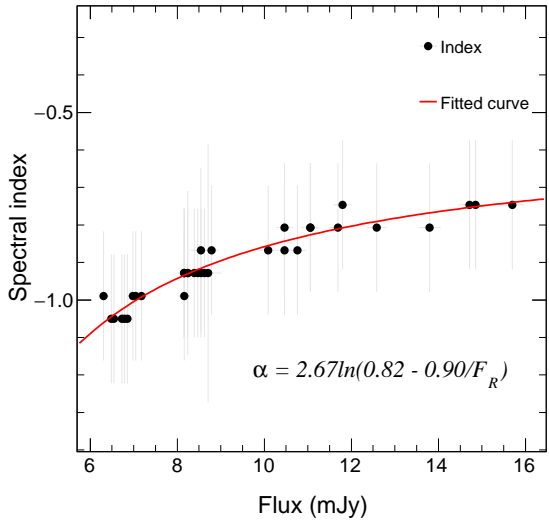}
    \caption{Same as Fig.~\ref{fig:spec-example}, but for BL Lac object 1ES 1959+650 during an outburst stage between MJD 55600 and MJD 56300.}
    \label{fig:spec-lnup}
\end{figure}

In general, the relationship between spectral indices and fluxes can be well fitted by equation~(\ref{Equ:fitln}). In order to assess the fitting quality, we have examined the fitted results source by source. On closer inspection, the spectral feature can be generally fitted well in both high and low states.
For instance, one can see that equation~(\ref{Equ:fitln}) is a good description of the spectral behavior of 1ES 1959 +650 whether the source is in a brighter or darker state, i.e., the flux is in the higher or lower part (see Fig.~\ref{fig:spec-example}). Only 3 sources, CTA 102, 3C 454.3 and OJ 287, can not be well fitted in their very-high states.
For each source, the spectral behaviors have been examined on different time scales.
They have not only been investigated on several-year time-scales, but also been examined on several-day to several-month time-scales.
Fig.~\ref{fig:spec-shortscale} shows an example of a 12-day short-term spectral behavior of B2 1633+382 during MJD 58281--58293.
The results demonstrate that the sources obey BSWB or RSWB trend on both long-term and short-term time-scales.
And furthermore, the sources have been checked during different stages, and the results show that the sources follow the law of BSWB or RSWB well whether during the outburst or descent stage. Fig.~\ref{fig:spec-lnup} exhibits an case of the spectral behavior of 1ES 1959+650 when it is in an outburst stage during MJD 55600--56300.

Several mechanisms have been proposed to interpret the phenomena of the spectral behavior. The two different spectral behaviors of BWB and RWB are explained by different models, and even for the same behavior, there exist different explanations. For the BWB trend, different models have been proposed, such as fresh electrons injection \citep{kirk98}, one-component synchrotron \citep{fiorucci04}, inter-band time delay \citep{wu07}, and two component contribution \citep{fiorucci04,ikejiri11}. Meanwhile, the RWB behavior has been interpreted as a strong contribution of a blue thermal emission from the accretion disc \citep{villata06,rani10,bonning12,isler17}, or changes in the Doppler factor \citep{raiteri17}.
\cite{zhang22} suggested that the BWB and RWB trends are special cases of the BSWB and RSWB trends in low states, respectively, and both of them arise from the same mechanism.
They proposed a new model with two constant-spectral-index components, and inferred a non-linear formula, which can well explain both the BSWB and RSWB trends, not only qualitatively but also quantitatively.

In this study, the fact that both BSWB and RSWB behaviors can be well fitted by the same non-linear formula (equation~(\ref{Equ:fitln})) implicates that both of them can be explained by the model of two constant-spectral-index components.
 The emission of optical band consists of a weak and relative stable thermal component, and a strong and highly variable synchrotron emission component, both of which have constant spectral indices. In the low state, the indices tend to be the thermal emission, while in the high state, since the emission begins to be dominated by the more variable non-thermal component, the index progressively approaches that of the non-thermal component.
 This mechanism naturally creates the BSWB or RSWB trend.

BL Lacs have less thermal emission contributions than FSRQs, therefore, their colors of thermal component are much redder than those of FSRQs.
And moreover, in most cases, for BL Lacs, the color of synchrotron emission component is bluer than that of the thermal component,
so most of BL Lacs exhibit BSWB trend.
While for FSRQs, the color of synchrotron emission component is redder than that of the thermal component, and as a result, most of them show RSWB trend.
However, there are several exceptions. For example, AO 0235+164, as a BL Lac object, however, is more like an FSRQ in many physical aspects \citep{chen01}, and it has a much more luminous accretion disc than other BL Lacs, even can be comparable to FSRQs \citep{ghisellini10,paliya21}, so it follows the RSWB trend.
The situation of the FSRQ 3C 279 is just the opposite, which implies that its thermal emission is relatively weak.
As another special case, the source FSRQ 3C 273 has a strong blue component from the accretion disc \citep{fernandes20,li20}, but it exhibits a slight BSWB trend, which may be due to its bluer synchrotron radiation.
Whether it is BL Lac object or FSRQ, its spectral behavior ultimately depends on the quantitative comparison between the index of two components, namely, the thermal emission component from the accretion disc and the synchrotron radiation component from the jet.

\section{Conclusions}
In sum, we have explored optical spectral features of 27 blazars, and confirmed that blazars follow the universal BSWB or RSWB trends.
The results suggest that FSRQs prefer the RSWB trend, while BL Lacs favor the BSWB trend.
These two trends can be described quantitatively by the same non-linear formula.
The model with two constant-spectral-index components can successfully interpret the optical spectral features both qualitatively and quantitatively, including the differences between BL Lac objects and FSRQs.

\section*{Acknowledgements}

We are grateful to the referee for his/her valuable suggestions.
This work has been supported by National Natural Science Foundation
 of China (Grant No. U1831124) and by Key Laboratory of Functional
 Materials and Devices for Informatics (Grant No. 2020XXGN01).
 We thanks Steward Observatory of University of Arizona for the long-term optical monitoring photometry data.

\section*{Data Availability}
All of the optical data used in this work are publicly available at the website http://james.as.arizona.edu/$\sim$psmith/Fermi.



\bibliographystyle{mnras}








\bsp	
\label{lastpage}
\end{document}